\documentclass[journal]{IEEEtran}

\usepackage[caption=true,font=normalsize,subrefformat=simple]{subfig}

\usepackage{url}
\usepackage{multirow}
\usepackage{booktabs}
\usepackage{color}
\usepackage{amsmath}
\usepackage{colortbl}
\usepackage{graphicx}
\usepackage{amssymb}

\usepackage{algorithm}
\usepackage{algpseudocode}

\usepackage{comment}

\usepackage{pifont}
\usepackage[table]{xcolor}

\usepackage{siunitx}
\usepackage{utfsym}
\usepackage{bm}

\usepackage{orcidlink}

\DeclareMathOperator{\softmax}{softmax}
\DeclareMathOperator{\sample}{sample}
\DeclareMathOperator{\cossim}{sim}
\DeclareMathOperator*{\topk}{top\,k}

\definecolor{ReviewColor}{HTML}{144eaf}

\definecolor{BaselineRowBg}{gray}{0.9}
\definecolor{MissingRowBg}{gray}{0.9}
\definecolor{SelectedRowBg}{HTML}{e1eafa}

\newcommand{\titlecaption}[2]{\caption{\textbf{#1}. #2}}

\begin{document}

\title{Self-Supervised Frameworks for Speaker Verification via Bootstrapped Positive Sampling}

\author{
    Theo~Lepage~\orcidlink{0009-0009-0676-4099},~\IEEEmembership{Student~Member,~IEEE,}
    and~Reda~Dehak~\orcidlink{0000-0002-4078-7261},~\IEEEmembership{Member,~IEEE}%
    \thanks{The authors are with EPITA Research Laboratory (LRE), France (e-mail: theo.lepage@epita.fr; reda.dehak@epita.fr).}%
    \thanks{Project source code and resources: \url{https://github.com/theolepage/sslsv}.}%
}


\maketitle

\begin{abstract}
Recent developments in Self-Supervised Learning (SSL) have demonstrated significant potential for Speaker Verification (SV), but closing the performance gap with supervised systems remains an ongoing challenge. SSL frameworks rely on anchor-positive pairs, constructed from segments of the same audio utterance. Hence, positives have channel characteristics similar to those of their corresponding anchors, even with extensive data-augmentation. Therefore, this positive sampling strategy is a fundamental limitation as it encodes too much information regarding the recording source in the learned representations. This article introduces Self-Supervised Positive Sampling (SSPS), a bootstrapped technique for sampling appropriate and diverse positives in SSL frameworks for SV. SSPS samples positives close to their anchor in the representation space, assuming that these pseudo-positives belong to the same speaker identity but correspond to different recording conditions. This method consistently demonstrates improvements in SV performance on VoxCeleb benchmarks when applied to major SSL frameworks, including SimCLR, SwAV, VICReg, and DINO. Using SSPS, SimCLR and DINO achieve 2.57\% and 2.53\% EER on VoxCeleb1-O, respectively. SimCLR yields a 58\% relative reduction in EER, getting comparable performance to DINO with a simpler training framework. Furthermore, SSPS lowers intra-class variance and reduces channel information in speaker representations while exhibiting greater robustness without data-augmentation.
\end{abstract}

\begin{IEEEkeywords}
Self-Supervised Learning, Speaker Recognition, Speaker Representations, Speech Processing.
\end{IEEEkeywords}

\section{Introduction}

\begin{figure}[t]
  \centering
  \subfloat[\textbf{Self-Supervised Learning (SSL)}\label{fig:ssps_overview_ssl}]{
    \includegraphics[width=0.8\linewidth]{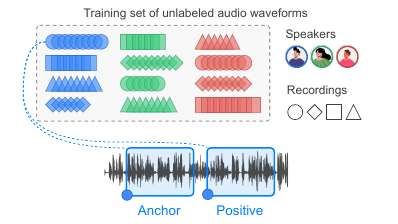}
  }
  \\
  \subfloat[\textbf{Self-Supervised Positive Sampling (SSPS)}\label{fig:ssps_overview_ssps}]{
    \includegraphics[width=0.8\linewidth]{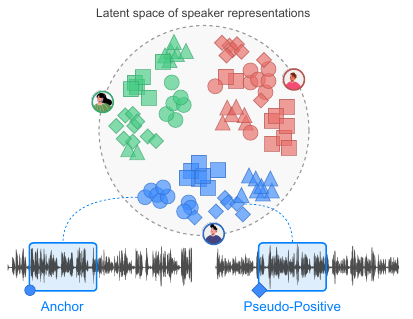}
  }
  \titlecaption{Overview of the positive sampling in SSL (a) and SSPS (b)}{Standard SSL samples a \textit{positive} from the same utterance as the \textit{anchor}, and thus from the \underline{same} recording. The proposed SSPS samples a \textit{pseudo-positive} from an utterance of a \underline{different} recording than the \textit{anchor} in the latent space, encouraging invariance to channel-related information.}
  \label{fig:ssps}
\end{figure}

\IEEEPARstart{S}{peaker} Recognition (SR) corresponds to the process of identifying the speaker's identity in an audio speech utterance. The main task in the SR field is Speaker Verification (SV), which aims to determine whether two speech utterances are spoken by the same speaker. To achieve this task, SR systems aim to define representations that maximize inter-speaker distances and minimize intra-speaker variances while being robust to extrinsic variabilities (e.g., noise, environment, recording device, and channel conditions).

With the emergence of deep learning, traditional machine learning methods, such as i-vectors \cite{dehak2011IVector}, have been surpassed by DNN models, including x-vectors \cite{snyder2018XVectors}, architectures based on ResNet \cite{chung2020DefenceMetricLearningSR}, and more recently, the ECAPA-TDNN \cite{desplanques2020ECAPATDNN} model. These methods are trained to match input speech utterances to their respective speaker identities in a supervised manner on a large corpus of annotated speech segments \cite{chung2018VoxCeleb2}. Indeed, the performance of deep learning methods scales with the amount of training data. However, this training paradigm represents a notable constraint due to the scarcity and expense of getting annotated training samples.

Self-Supervised Learning (SSL) methods have emerged as a solution to this bottleneck by learning meaningful representations from the input data. SSL has been shown to enhance the scalability potential of models by leveraging the abundance of unlabeled data available. Various SSL frameworks have been proposed in computer vision following the joint-embedding architecture, which considers an \textit{anchor} and a \textit{positive} sample derived from different augmentations of the same input, thereby representing the same high-level information. Contrastive learning, adopted by SimCLR \cite{chen2020SimCLR} and MoCo \cite{he2020MoCo}, aims to maximize the similarity of positive pairs while minimizing the similarity of negative pairs sampled from the current batch or a memory queue. The effect of class collision, occurring when randomly sampled negatives belong to the same class as the anchor-positive pair, is considered negligible on the training convergence. Clustering-based methods, such as DeepCluster \cite{caron2018DeepCluster} and SwAV \cite{caron2020SwAV}, learn by grouping similar representations into clusters and using the resulting assignments as a supervisory signal. Information maximization techniques like Barlow Twins \cite{zbontar2021BarlowTwins} and VICReg \cite{bardes2022VICReg} apply constraints on the learned embedding space to enforce invariance between the anchor and the positive representations while explicitly avoiding collapse (i.e., non-informative representations). Finally, methods based on self-distillation, such as BYOL \cite{grill2020BYOL} and DINO \cite{caron2021DINO}, leverage knowledge distillation, where a student model is trained to match the output of a teacher model.

These approaches have also been successfully extended to the downstream task of SV. Most methods are based on contrastive learning \cite{huh2020AAT,zhang2021SimCLR,xia2021MoCo,lepage2024AdditiveMargin} and self-distillation, as the DINO framework currently achieves state-of-the-art performance on SV \cite{zhang2022C3DINO,chen2023RDINO,han2024CADINO,cho2022DINO,heo2022DINOCurriculum}. In this context, SSL relies on the assumption that the anchor and the positive belong to the same speaker, as both segments are extracted from the same speech utterance. Moreover, extensive data-augmentation is applied to both inputs to learn representations robust against extrinsic variabilities (e.g., environmental noise, reverberation, acoustic conditions, recording devices).

Nonetheless, data-augmentation is insufficient to overcome the effect of SSL same-utterance positive sampling, which encodes various channel characteristics in speaker representations, yielding high variance within same-speaker representations. Several techniques have been explored to address this issue in SV: AP+AAT \cite{huh2020AAT} introduces an adversarial loss to penalize the model from learning channel information; i-mix \cite{kang2022LMix} is a data-driven augmentation strategy that interpolates training utterances to make the model focus on the main distinguishing factor between them; DPP \cite{tao2023DPP} finds diverse positive samples using another modality by cross-referencing speech and face data; CA-DINO \cite{han2024CADINO} clusters the embeddings to sample a positive from the same speaker class as the anchor. Finally, alternative positive sampling strategies have been proposed in computer vision, such as NNCLR \cite{dwibedi2021NNCLR} and GPS-SSL \cite{feizi2024GPSSSL}, which select positives in the latent space through nearest neighbor search.

A new positive sampling strategy for SSL frameworks, named \textbf{\underline{S}elf-\underline{S}upervised \underline{P}ositive \underline{S}ampling (SSPS)}, is introduced in this article. Instead of sampling a positive from the same utterance as the anchor (Figure~\subref*{fig:ssps_overview_ssl}), this bootstrapped approach determines a \textit{pseudo-positive} from a different utterance by relying on the knowledge acquired and refined by the SSL model (Figure~\subref*{fig:ssps_overview_ssps}). After several epochs of standard SSL training, it is assumed that utterances of the same speaker identity with different recording conditions will have representations close to the anchor's representation in the learned speaker representation space. Different sampling algorithms based on nearest neighbors (SSPS-NN) and clustering (SSPS-Clustering) are presented. This method enables the learning of more robust representations by matching different recordings to the same speaker identity. SSPS significantly improves the performance on SV for all major SSL frameworks by reducing the intra-class variance of speaker representations. Additionally, a decrease in the information related to the source recording is observed in the learned representations when using SSPS. Furthermore, experiments show that this method is less dependent on data-augmentation as pseudo-positives are sampled from different recordings with varying channel information. This manuscript is an extended version of a previously published conference article \cite{lepage2025SSPS}, broadening the scope to a wider range of SSL frameworks, introducing two distinct sampling algorithms, and providing deeper insights into the effects of SSPS on the learned speaker representations.

The general SSL training scheme and major frameworks are described in Section~\ref{sec:sslsv}. Then, SSPS and its instantiations, SSPS-NN and SSPS-Clustering, are introduced in Section~\ref{sec:ssps}. The experimental setup is detailed in Section~\ref{sec:exp_setup}. Subsequently, Section~\ref{sec:results} highlights the limitations of SSL positive sampling and presents preliminary results obtained with SSPS. It then evaluates SSPS on SV, compares it with state-of-the-art methods, and analyzes the resulting learned representations. Finally, the article concludes in Section~\ref{sec:conclusions}.
\begin{figure*}
  \centering
  \includegraphics[width=\textwidth]{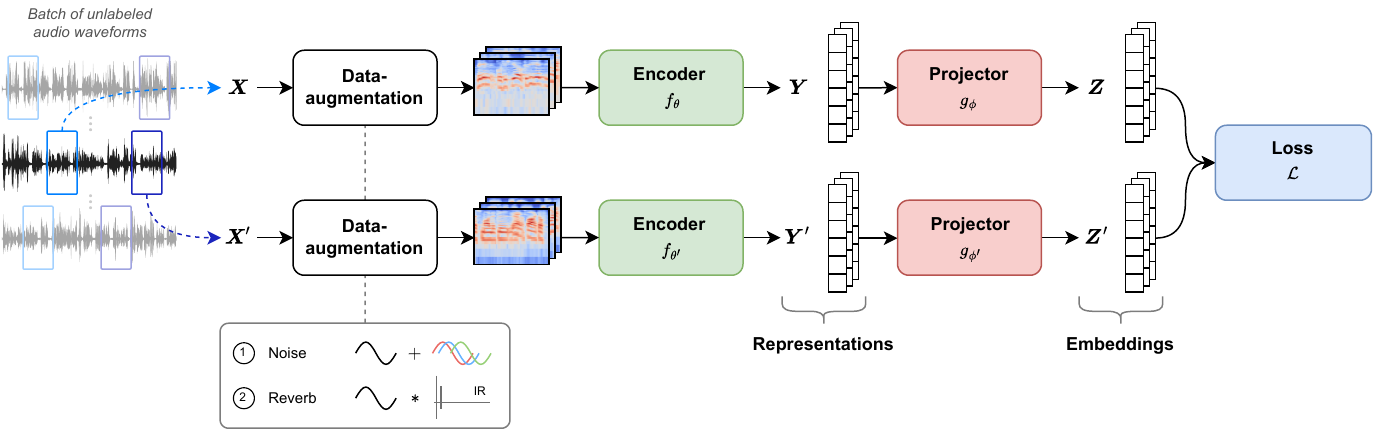}
  \titlecaption{SSL training framework for SV}{The joint-embedding architecture generates a pair of representations (for the downstream task) and a pair of embeddings (for the pretext task) from a given unlabeled audio sample.}
  \label{fig:SSL_SV_training_framework}
\end{figure*}

\section{Self-Supervised Learning Frameworks for Speaker Verification}
\label{sec:sslsv}

The self-supervised training framework is based on the \textit{symmetrical} or \textit{asymmetrical} joint-embedding architecture to produce a pair of embeddings from a given unlabeled audio waveform, as illustrated in Figure~\ref{fig:SSL_SV_training_framework}.

Let $\mathcal{I} \equiv\{1, \ldots, N\}$ be the indices of the training set composed of $N$ samples. At each training iteration, $B$ utterances are sampled such that $\mathcal{B} \subseteq \mathcal{I}$ are the mini-batch indices. For each utterance $u_i \in \{u_i\}_{i \in \mathcal{B}}$, two segments, $\boldsymbol{x}_i$ and $\boldsymbol{x}_i^{\prime}$, are randomly extracted. They represent the \textit{anchor} and the \textit{positive}, respectively. Next, random data-augmentation is applied to each segment, and their mel-scaled spectrogram features are used as input features.

The joint-embedding architecture consists of two branches, one with an encoder $f_{\theta}$ and projector $g_{\phi}$, and the other with an encoder $f_{\theta^{\prime}}$ and projector $g_{\phi^{\prime}}$, where the corresponding modules share the same architecture across branches. First, encoders $f_{\theta}$ and $f_{\theta^{\prime}}$ transform $\boldsymbol{x}_i$ and $\boldsymbol{x}_i^{\prime}$ to their representations $\boldsymbol{y}_i$ and $\boldsymbol{y}_i^{\prime}$ of dimension $D_{\text{repr}}$. Then, projectors $g_{\phi}$ and $g_{\phi^{\prime}}$ map $\boldsymbol{y}_i$ and $\boldsymbol{y}_i^{\prime}$ to their corresponding embeddings $\boldsymbol{z}_i$ and $\boldsymbol{z}_i^{\prime}$ of dimension $D_{\text{emb}}$. 

Representations serve as speaker features to perform SV, while embeddings are used to compute the loss $\mathcal{L}$ and optimize the model. For the downstream task of SV, SSL methods aim to minimize the distance between same-speaker embeddings while avoiding collapse (i.e., trivial solutions leading to non-informative representations). In this context, using distinct segments and applying data-augmentation is fundamental to prevent collapse and to produce robust representations that primarily capture speaker identity.

In the default training scheme, SSL frameworks adopt the \textit{symmetrical} joint-embedding architecture (e.g., SimCLR, SwAV, and VICReg), where the module weights are shared such that $\theta^{\prime} \gets \theta$ and $\phi^{\prime} \gets \phi$. When employing the \textit{asymmetrical} version (e.g., MoCo and DINO), one branch is referred to as the \textit{student} or \textit{query}, and the other branch is referred to as the \textit{teacher} or \textit{key}. In this particular case, the gradient is not propagated through the teacher branch and the teacher weights are updated with an Exponential Moving Average (EMA) of the student weights such that $\theta^{\prime} \gets m\theta^{\prime} + (1-m)\theta$ and $\phi^{\prime} \gets m\phi^{\prime} + (1-m)\phi$ where $m \in [0, 1)$ is the momentum update coefficient.

As training processes mini-batches, the following notations are introduced to represent input segments, representations, and embeddings: $\boldsymbol{X}=\{\boldsymbol{x}_i\}_{i \in \mathcal{B}}$, $\boldsymbol{X}^{\prime}=\{\boldsymbol{x}^{\prime}_i\}_{i \in \mathcal{B}}$, $\boldsymbol{Y}=\{\boldsymbol{y}_i\}_{i \in \mathcal{B}}$, $\boldsymbol{Y}^{\prime}=\{\boldsymbol{y}_i^{\prime}\}_{i \in \mathcal{B}}$, $\boldsymbol{Z}=\{\boldsymbol{z}_i\}_{i \in \mathcal{B}}$, and $\boldsymbol{Z}^{\prime}=\{\boldsymbol{z}_i^{\prime}\}_{i \in \mathcal{B}}$. $\boldsymbol{z}_i$ denotes the embedding of the $i$-th sample from $\boldsymbol{Z}$ and $\boldsymbol{z}^d$ denotes the $d$-th dimension of all embeddings in $\boldsymbol{Z}$.

The following presents major SSL frameworks from different paradigms. SimCLR and MoCo, based on contrastive learning, are described in Sections \ref{sec:simclr} and \ref{sec:moco}, respectively. Then, SwAV, a clustering-based approach, is introduced in Section~\ref{sec:swav}. Next, VICReg, following the information maximization paradigm, is presented in Section~\ref{sec:vicreg}. Finally, DINO, based on self-distillation, is described in Section~\ref{sec:dino}. This set of methods allows the majority of SSL-based frameworks to be covered.

The similarity between two embeddings $\boldsymbol{u}$ and $\boldsymbol{v}$ is defined as $\ell(\boldsymbol{u}, \boldsymbol{v})= \exp\left(\cossim\left(\boldsymbol{u}, \boldsymbol{v}\right) / \tau\right)$ where $\tau$ is a temperature scaling hyperparameter. $\cossim(\boldsymbol{u}, \boldsymbol{v})$ corresponds to the cosine similarity and is obtained by computing the dot product between the two $l_2$ normalized embeddings $\boldsymbol{u}$ and $\boldsymbol{v}$.

\subsection{SimCLR}
\label{sec:simclr}
SimCLR \cite{chen2020SimCLR, chen2020SimCLRV2} implements the concept of contrastive learning, which aims to maximize the similarity between the anchor and the positive, while minimizing the similarity between the anchor and the negative samples. Negatives are randomly selected from the current mini-batch, assuming that these utterances belong to a different speaker identity.

The objective function $\mathcal{L}_{\text {SimCLR}}$, which is equivalent to the NT-Xent loss derived from InfoNCE \cite{vandenoord2019CPC,rusak2025InfoNCE}, is defined as:
\begin{equation}
    \label{eq:simclr_loss}
    \mathcal{L}_{\text{SimCLR}} = - \frac{1}{B} \sum_{i \in \mathcal{B}} \log \frac{\ell \left(\boldsymbol{z}_i, \boldsymbol{z}_i^{\prime} \right)}{\sum_{j \in \mathcal{B}} \ell \left(\boldsymbol{z}_i, \boldsymbol{z}_j^{\prime} \right)}.
\end{equation}

\subsection{MoCo}
\label{sec:moco}
MoCo \cite{he2020MoCo, chen2020MoCoV2} follows the contrastive learning paradigm and employs the asymmetrical joint-embedding architecture, implying that the key weights are updated with an EMA of the query weights. A queue $\boldsymbol{Q}^{\text{MoCo}}$ of recent key embeddings, dynamically updated at each training step, is introduced to increase the number of negatives.

The objective function $\mathcal{L}_{\text {MoCo}}$ is defined as:
\begin{equation}
    \label{eq:moco_loss}
    \mathcal{L}_{\text {MoCo}}=- \frac{1}{B} \\ \sum_{i \in \mathcal{B}} \log \frac{\ell \left(\boldsymbol{z}_i, \boldsymbol{z}_i^{\prime}\right)}{\ell \left(\boldsymbol{z}_{i} , \boldsymbol{z}_i^{\prime} \right) + \sum\limits_{j \in \mathcal{J}} \ell \left(\boldsymbol{z}_{i} , \boldsymbol{q}^{\text{MoCo}}_j \right)},
\end{equation}
where $\mathcal{J} \equiv\{1, \ldots, |\boldsymbol{Q}^{\text{MoCo}}|\}$ and $\boldsymbol{q}^{\text{MoCo}}_i$ is the $i$-th element of the queue $\boldsymbol{Q}^{\text{MoCo}}$.

\subsection{SwAV}
\label{sec:swav}
SwAV \cite{caron2020SwAV} simultaneously clusters embeddings while enforcing consistency between assignments produced for the anchor and the positive. Prototypes are learned jointly with the model and soft assignments are generated online using the Sinkhorn-Knopp algorithm \cite{cuturi2013Sinkhorn}. Following a ``swapped" prediction mechanism, the code of the anchor is predicted from the embedding of the positive, and vice versa. A queue $\boldsymbol{Q}^{\text{SwAV}}$ of latest embeddings is used to simulate a larger batch size, thereby providing more stable cluster assignments.

The objective function $\mathcal{L}_{\text{SwAV}}$ is defined as:
\begin{equation}
    \label{eq:swav_loss}
    \mathcal{L}_{\text {SwAV}} = - \frac{1}{B \cdot P} \sum_{i \in \mathcal{B}} \sum_{k=1}^{P} \boldsymbol{a}_{i,k} \log \frac{\ell \left(\boldsymbol{z}_i^{\prime}, \boldsymbol{p}_{k}\right)}{\sum\limits_{k^{\prime}=1}^{P} \ell \left(\boldsymbol{z}_{i}^{\prime} , \boldsymbol{p}_{k^{\prime}} \right)},
\end{equation}
where $P$ is the number of prototypes, $\boldsymbol{p}_k$ is the $k$-th prototype, and $\boldsymbol{a}_{i, k}$ is the code of $\boldsymbol{z}_i$ for the $k$-th prototype.

\subsection{VICReg}
\label{sec:vicreg}
VICReg \cite{bardes2022VICReg} maximizes information directly from the embeddings and provides an explicit solution to avoid collapse. The \textit{variance} ($v$) component enforces the variance to reach $1$ along the $D_{\text{emb}}$ dimensions of the embeddings to produce class-dependent representations and prevent a collapsing solution, such that $v\left(\boldsymbol{Z}\right)=\frac{1}{D_{\text{emb}}} \sum_{d=1}^{D_{\text{emb}}} \max \left(0, 1-\sqrt{\operatorname{Var}(\boldsymbol{z}^d)}\right)$. The \textit{invariance} ($s$) component minimizes the distance between the anchor and positive to learn invariance to perturbations created through data-augmentation, such that $s\left(\boldsymbol{Z}, \boldsymbol{Z}^{\prime}\right)=\frac{1}{B} \sum_{i \in \mathcal{B}}\left\|\boldsymbol{z}_{i}-\boldsymbol{z}_{i}^{\prime}\right\|_{2}^{2}$. The \textit{covariance} ($c$) component makes the off-diagonal coefficients of the embeddings covariance matrix $C$ to be close to $0$ to prevent the dimensions of the embeddings from encoding similar information, such that $c\left(\boldsymbol{Z}\right)=\frac{1}{D_{\text{emb}}} \sum_{d=1}^{D_{\text{emb}}} \sum_{\substack{d^{\prime}=1 \\ d^{\prime} \neq d}}^{D_{\text{emb}}} [C(\boldsymbol{Z})]_{d, d^{\prime}}^{2}$.

The objective function $\mathcal{L}_{\text{VICReg}}$ is defined as:
\begin{align}
    \label{eq:vicreg_loss}
    \mathcal{L}_{\text{VICReg}} 
    &= \lambda \, s\left(\boldsymbol{Z}, \boldsymbol{Z}^{\prime}\right) \notag \\
    &\quad + \mu\left(v(\boldsymbol{Z}) + v\left(\boldsymbol{Z}^{\prime}\right)\right) \notag \\
    &\quad + \nu\left(c(\boldsymbol{Z}) + c\left(\boldsymbol{Z}^{\prime}\right)\right),
\end{align}
where $\lambda$, $\mu$ and $\nu$ are hyperparameters to scale the invariance, variance and covariance terms.

\subsection{DINO}
\label{sec:dino}
DINO \cite{caron2021DINO} is a self-distillation framework based on the knowledge distillation paradigm where a student model is trained to predict the output distribution of a teacher model. Following the ``multi-crop'' strategy \cite{caron2020SwAV}, a larger set of augmented utterances with different segment lengths is considered, resulting in four small (\textit{local}) and two large (\textit{global}) segments. All views are fed through the student, while only global views are fed through the teacher to learn ``local-to-global'' correspondences. The teacher is updated with an EMA of the student weights following the asymmetrical joint-embedding architecture. To avoid collapse, \textit{sharpening} and \textit{centering} are applied to the teacher outputs: centering prevents one dimension from prevailing while sharpening discourages collapse to the uniform distribution.

The objective function $\mathcal{L}_{\text{DINO}}$ is defined as:
\begin{equation}
    \label{eq:dino_loss}
    \mathcal{L}_{\text{DINO}} =
        \frac{1}{B} \sum_{i \in \mathcal{B}}
        \sum_{t=1}^{2} \sum_{\substack{s=1 \\ s \neq t}}^{6}
        H\bigg(
            \frac{\boldsymbol{z}_{i,t}^{\prime} - \boldsymbol{c}}{\tau_{\text{t}}},
            \frac{\boldsymbol{z}_{i,s}}{\tau_{\text{s}}}
        \bigg),
\end{equation}
where $H\left(\boldsymbol{a}, \boldsymbol{b}\right)=- \softmax(\boldsymbol{a}) \log \left( \softmax(\boldsymbol{b}) \right)$, $\boldsymbol{z}_{i,t}^{\prime}$ is the $t$-th teacher embedding of the $i$-th sample, $\boldsymbol{z}_{i,s}$ is the $s$-th student embedding of the $i$-th sample, $\tau_{\text{t}}$ is the teacher temperature, $\tau_{\text{s}}$ is the student temperature, and $\boldsymbol{c}$ is a running mean of the teacher embeddings.

\begin{figure*}
  \centering
  \includegraphics[width=\textwidth]{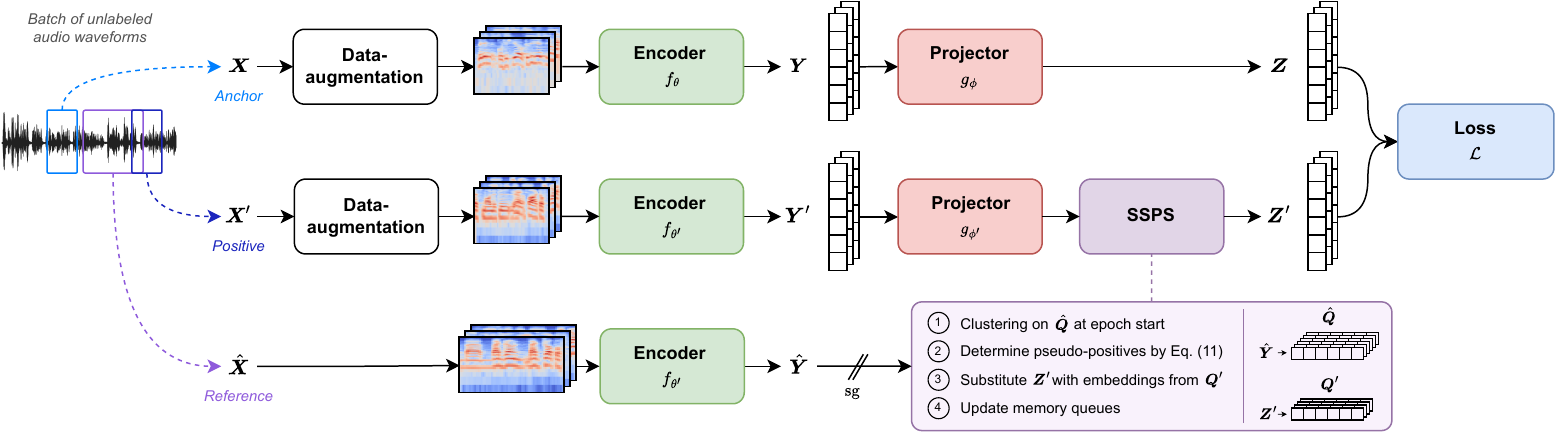}
  \titlecaption{SSL training framework for SV with SSPS-Clustering}{Compared to the standard framework, the \textit{positive} is substituted by a \textit{pseudo-positive}, which is retrieved from a memory queue based on the \textit{anchor}'s clustering assignment. Clustering is performed at the beginning of each epoch on \textit{reference} representations derived from longer and non-augmented audio segments.}
  \label{fig:SSPS_SSL_SV_training_framework}
\end{figure*}
\section{SSPS: Self-Supervised Positive Sampling}
\label{sec:ssps}

Different studies have demonstrated theoretically that the primary factors influencing the quality of SSL representations lie in how the positives are defined \cite{balestriero2023CookbookSSL}. This incorporated ``a priori'' knowledge helps model class distribution, which is fundamental to learning meaningful representations without labels and human supervision.

This section introduces the proposed positive sampling approach, presenting its motivation (Section~\ref{sec:ssps-motivation}), the underlying assumptions and properties (Section~\ref{sec:ssps-method}), the overall sampling method and its integration into the SSL framework (Section~\ref{sec:ssps-framework}), and two instantiations of the strategy: SSPS-NN (Section~\ref{sec:ssps-nn}) and SSPS-Clustering (Section~\ref{sec:ssps-clustering}).

\subsection{Motivation}
\label{sec:ssps-motivation}

SSL models trained for SV are prone to learning channel-related information, as anchor-positive pairs are constructed from segments of the same audio utterance, thereby from the same recording. This effect is amplified by training datasets collected "in the wild", which contain speech segments corrupted with various channel effects and other factors.

The standard strategy to overcome this bias is to rely on data-augmentation techniques. In SR, the challenge is designing transformations that replicate various acoustic conditions to prevent the model from relying on channel characteristics. However, conventional audio augmentation techniques, such as adding background noise and reverberation, fall short of limiting the impact of the same-utterance positive sampling. These methods do not appear to be sufficient for simulating the diversity among samples from a given speaker class. Moreover, data-augmentation has inherent limitations since it highly depends on the input data and the downstream task.

Therefore, SSL frameworks lack the ability to match two utterances with different acoustic conditions to the same speaker identity. This represents the primary bottleneck to achieving the performance of supervised systems, which inherently address this challenge using human-annotated labels.

\subsection{Method}
\label{sec:ssps-method}
\textbf{\underline{S}elf-\underline{S}upervised \underline{P}ositive \underline{S}ampling (SSPS)} is proposed to sample more relevant positives, denoted as \textit{pseudo-positives}, using the acquired knowledge of the SSL model.

SR datasets, such as VoxCeleb, consist of multiple recordings (or sessions) representing several utterances from a single speaker, and each speaker has multiple recordings. Based on this observation and the fact that SSL representations encode channel information, the following assumption is formulated: ``SSL same-utterance positive sampling group representations of the same recordings, sharing similar channel information, before modeling speaker identities''.

Suppose that representations $\boldsymbol{y}_1$ is from speaker $S$ and recording $R$, $\boldsymbol{y}_2$ is from the same recording but a distinct utterance, $\boldsymbol{y}_3$ is from the same speaker but a different recording, and $\boldsymbol{y}_4$ is from a different speaker. In this case, the assumption can be expressed as $\cossim\left(\boldsymbol{y}_1,\boldsymbol{y}_2\right) \geq \cossim\left(\boldsymbol{y}_1,\boldsymbol{y}_3\right) \geq \cossim\left(\boldsymbol{y}_1,\boldsymbol{y}_4\right)$, where $\cossim(\boldsymbol{u}, \boldsymbol{v})$ is the cosine similarity between two vectors. This suggests that the latent space is first organized by groups of utterances from the same recording and then by groups of utterances from the same speaker.

SSPS aims to sample accurate pseudo-positives from the same speakers but from different recordings. This approach falls within the family of bootstrapped techniques, as the selection of pseudo-positives relies entirely on the model's own evolving representations, without using any labels, and improves over iterations (i.e., enhanced representations lead to better sampling and vice versa). This method can be implemented into any SSL framework and introduces a novel axis of freedom, namely, the design of the positive sampling algorithm. It extends SSL beyond tweaking the architecture of the model, the data-augmentation, or the objective function.

Positive sampling and data-augmentation are complementary strategies that serve the common goal of generating meaningful anchor-positive pairs to shape the underlying speaker distribution in the latent space. Data-augmentation is crucial when employing the standard SSL same-utterance positive sampling, as both the anchor and the positive originate from the same recording and would otherwise have very similar acoustic conditions. However, since SSPS creates anchor-positive pairs from different recordings of the same speaker, it inherently introduces greater variability. It thus has the notable property of being less reliant on data-augmentation, particularly when the training data exhibits diverse acoustic conditions. This property is demonstrated through experiments presented at the end of the article.

\subsection{Framework}
\label{sec:ssps-framework}
Given a training utterance $u_i$ with $i \in \mathcal{B}$, from which the anchor is sampled, let $pos(i)$ denote the index of an utterance $u_{pos(i)}$ used to sample the positive. Standard SSL creates the positive from the same utterance by setting $pos(i)=i$ (Figure~\subref*{fig:ssps_overview_ssl}) and applying random data-augmentation. In contrast, SSPS selects a pseudo-positive from a different utterance in latent space, such that $pos(i) \neq i$ (Figure~\subref*{fig:ssps_overview_ssps}).

The proposed framework is built around the following main components: (1) reference representations, extracted from longer unaltered audio segments to obtain a robust latent space of training samples; (2) two memory queues, storing all reference representations and latest positive embeddings throughout training; and (3) pseudo-positive sampling, which determines $pos(i)$ based on the latent space of reference representations and retrieves the corresponding positive embedding from the memory queue.

\subsubsection{Reference representations}
Since SSPS aims to capture the underlying data patterns for identifying speakers and recordings, the anchor and positive representations, $\boldsymbol{y}_i$ and $\boldsymbol{y}_i^{\prime}$, are not used to determine $pos(i)$. Instead, a reference segment $\hat{\boldsymbol{x}_i}$, extracted from $u_i$ but without data-augmentation and using a longer audio segment, is introduced. As shown in Fig.~\ref{fig:SSPS_SSL_SV_training_framework}, the corresponding reference representation $\hat{\boldsymbol{y}_i}$ is computed during training and used by the SSPS sampling algorithm.

\subsubsection{Memory queues}
\newcommand{\Qpos}{\boldsymbol{Q}^{\prime}}
\newcommand{\qpos}{\boldsymbol{q}^{\prime}}
\newcommand{\Qref}{\hat{\boldsymbol{Q}}}
\newcommand{\qref}{\hat{\boldsymbol{q}}}
SSPS relies on two memory queues, $\Qref$ of size $(|\Qref|, D_{\text{repr}})$ and $\Qpos$ of size $(|\Qpos|, D_{\text{emb}})$. $\Qref$ is used to store the reference representations $\{\hat{\boldsymbol{y}}_i\}_{i \in \mathcal{I}}$ and $\Qpos$ is used to store the positive embeddings $\{\boldsymbol{z}^{\prime}_i\}_{i \in \mathcal{I}}$. The former helps the sampling algorithm to determine which samples to use as pseudo-positives (i.e., determining $pos(i)$), while the latter is used to extract the corresponding positive embeddings for the SSL objective function.
By default, $|\Qref|$ and $|\Qpos|$ are set to $N$ to cover all training samples, but $|\Qpos|$ can be reduced depending on the setup to limit memory usage.
Memory queues are updated dynamically at each training iteration using $\hat{\boldsymbol{Y}}$ and $\boldsymbol{Z}^{\prime}$, respectively, as illustrated in Fig.~\ref{fig:SSPS_SSL_SV_training_framework}. For clarity, queues are indexed by training index, meaning $\qref_i$ and $\qpos_i$ retrieve the element corresponding to the $i$-th training sample, not the $i$-th position in the queue. 

\subsubsection{Pseudo-positive sampling}
After a pre-defined number of training epochs with the standard SSL framework, it is assumed that same-speaker representations are grouped in the latent space but with a large scatter due to the strong representation of channel and recording conditions. This assumption is used to sample a pseudo-positive from the neighborhood of the anchor that does not share the same channel and recording conditions. pseudo-positive embeddings will be sampled from $\Qpos$ if and only if $pos(i) \neq \varnothing$, otherwise, default SSL positive sampling is applied. For most SSL frameworks, the positive embedding of the anchor-positive pair of the $i$-th sample will be $\qpos_{pos(i)}$ extracted from $\Qpos$, instead of $\boldsymbol{z}_i^{\prime}$ from the mini-batch, in the objective functions $\mathcal{L}_{\text{SimCLR}}$ (\ref{eq:simclr_loss}), $\mathcal{L}_{\text{MoCo}}$ (\ref{eq:moco_loss}), $\mathcal{L}_{\text{SwAV}}$ (\ref{eq:swav_loss}), and $\mathcal{L}_{\text{VICReg}}$ (\ref{eq:vicreg_loss}). For the DINO framework, student outputs are considered to be anchors and teacher outputs to be positives, which implies that the teacher embeddings of the $i$-th sample will be $\qpos_{pos(i),1}$ and $\qpos_{pos(i),2}$, instead of $\boldsymbol{z}_{i,1}^{\prime}$ and $\boldsymbol{z}_{i,2}^{\prime}$ from the mini-batch, in the objective function $\mathcal{L}_{\text{DINO}}$ (\ref{eq:dino_loss}).

Note that SSPS can be made symmetric by using $\boldsymbol{q}_{pos(i)}$, extracted from a new memory queue $\boldsymbol{Q}$ similar to $\Qpos$, as a pseudo-positive for $\boldsymbol{z}_i^{\prime}$. However, this technique was not considered as preliminary experiments demonstrated that it results in equivalent downstream performance.

\subsection{SSPS-NN: Nearest Neighbors sampling}
\label{sec:ssps-nn}
Following the k-nearest neighbors (k-nn) algorithm, samples that are close to an utterance in the latent space can be considered as appropriate pseudo-positives.

The hyperparameter $M$ denotes the sampling window (i.e., the number of nearest neighbors considered for sampling).

Let $\mathcal{N}_i$ be the set of training indices that correspond to the $M$ nearest reference representations from the $i$-th sample:

\begin{equation}
    \mathcal{N}_i \triangleq \topk_{
        j \neq i
    } \big( \left\{\cossim \left(\qref_i, \qref_j\right), \forall j \in \mathcal{I} \right\} ; M \big),
\end{equation}
where $\topk \left(S; M\right)$ corresponds to the indices of the highest $M$ values from a set $S$, sorted in descending order to have the largest cosine similarities first.

The proposed sampling, denoted \textbf{SSPS-NN}, selects a pseudo-positive for the $i$-th sample such that:
\begin{equation}
    pos(i) = \sample \left( \mathcal{N}_{i} \right),
\end{equation}
where $\sample(S)$ corresponds to a random selection from a set $S$ using a uniform probability distribution.

Note that SSPS fallbacks to default SSL positive sampling if $\qpos_{pos(i)}$ is not present in $\Qpos$ (occurring when $|\Qpos| < N$) by setting $pos(i)=\varnothing$.

This approach is similar to NNCLR \cite{dwibedi2021NNCLR} and GPS-SSL \cite{feizi2024GPSSSL} but is limited in the context of SR because pseudo-positives will most likely belong to the same recording as their anchor, even if the sampling window $M$ is large.

\subsection{SSPS-Clustering: Clustering-based sampling}
\label{sec:ssps-clustering}
Clustering methods can also be employed to sample pseudo-positives from the same class or a neighboring class, given the clustering assignment of an utterance.

The hyperparameter $K$ denotes the number of clusters and $M$ controls the cluster sampling strategy: same-cluster sampling of Eq.~(\ref{eq:c}) is used when $M=0$ and neighboring-clusters sampling of Eq.~(\ref{eq:c_hat}) is used when $M>0$.

The k-means algorithm is applied to the reference representations from $\Qref$ to group the $|\Qref|$ utterances into $K$ clusters. The clustering is performed at the beginning of each training epoch to refine assignments as the quality of SSL representations improves. Let $c_i$ be the assigned cluster of the $i$-th utterance, and $\boldsymbol{m}_k$ the centroid of the $k$-th cluster.

\subsubsection{Same-cluster sampling}
Samples belonging to the same cluster as an utterance can be used to select the corresponding pseudo-positives because the clustering groups speaker identities when using an appropriate number of clusters, i.e., the hyperparameter $K$ should tend to the actual number of speakers in the training set. Therefore, the cluster $\hat{c}_i$ from which the pseudo-positive of the $i$-th sample will be selected can be set to the assigned cluster of this utterance, such that:
\begin{equation}
    \label{eq:c}
    \hat{c}_i = c_i.
\end{equation}
The disadvantage of this approach is that sampled pseudo-positives may originate from the same recording as the anchor and share similar channel information.

\subsubsection{Neighboring-clusters sampling}
Based on the assumption regarding the modeling of recording information before speaker identities, samples from neighboring clusters can be considered as pseudo-positives when choosing a number of clusters $K$ that is significantly larger than the number of speakers and tends to the number of recordings in the training set. The purpose is to select a class close enough to $c_i$ in the latent space to maximize the probability of being from the same speaker while having different channel characteristics. Hence, the sampling cluster $\hat{c}_i$ can be defined as:
\begin{equation}
    \label{eq:c_hat}
    \hat{c}_i = \sample \left( \mathcal{C}_{c_i} \right),
\end{equation}
where $\mathcal{C}_k$ is the set of $M$ nearest clusters from the $k$-th cluster in the latent space and is defined as:
\begin{equation}
    \label{eq:C}
    \mathcal{C}_k \triangleq \topk_{j \neq k} \big( \{ \cossim \left( \boldsymbol{m}_k, \boldsymbol{m}_j \right), \forall j \in [1, K] \} ; M \big).
\end{equation}

The proposed sampling, denoted \textbf{SSPS-Clustering}, selects a pseudo-positive for the $i$-th sample by randomly selecting a sample assigned to the cluster $\hat{c}_i$, such that:
\begin{equation}
    \label{eq:pos_i}
    pos(i) = \sample \left( \mathcal{S}_{\hat{c}_i} \right),
\end{equation}
where the set $\mathcal{S}_k$ contains the indices of training samples assigned to the $k$-th cluster and is defined as:
\begin{equation}
    \label{eq:S}
    \mathcal{S}_k \triangleq \{ j \in \mathcal{I} \text{ s.t. } c_j = k \}.
\end{equation}

Note that SSPS fallbacks to default SSL positive sampling if $\qpos_{pos(i)}$ is not present in $\Qpos$ (occurring when $|\Qpos| < N$) by setting $pos(i)=\varnothing$.

The preliminary experiments also include \textbf{SSPS-Clustering~(C)}, which corresponds to using $\boldsymbol{m}_{\hat{c}_i}$ (i.e., the centroid of the sampling cluster $\hat{c}_i$) instead of $\qpos_{pos(i)}$ as the pseudo-positive embedding.

For an overview of the clustering-based approach, please refer to the pseudo-code in Algorithm~\ref{alg:ssps_init} for epoch initialization and Algorithm~\ref{alg:ssps_iter} for training iterations.

\begin{algorithm}
\caption{SSPS-Clustering: Epoch Initialization}
\label{alg:ssps_init}
\begin{algorithmic}[1]
    \State \textbf{Inputs:} $K$, $M$, $\Qref$

    \State Run k-means on $\Qref$ with $K$ clusters
    \State Determine $\{\mathcal{C}_{k}\}_{k=1,\dots,K}$ by Eq.~(\ref{eq:C})
    \State Determine $\{\mathcal{S}_{k}\}_{k=1,\dots,K}$ by Eq.~(\ref{eq:S})

    \State \textbf{Return} $\{\mathcal{C}_{k}\}$, $\{\mathcal{S}_{k}\}$
\end{algorithmic}
\end{algorithm}

\begin{algorithm}
\caption{SSPS-Clustering: Training Iteration}
\label{alg:ssps_iter}
\begin{algorithmic}[1]
    \State \textbf{Inputs:} $\mathcal{B}$, $\hat{\boldsymbol{Y}}$, $\boldsymbol{Z}$, $\boldsymbol{Z}^{\prime}$, $\mathcal{L}$, $\Qpos$

    \State $\boldsymbol{Z}_{\text{SSPS}}^{\prime} \gets \left\{ 
    \begin{array}{ll}
        \qpos_{pos(i)}, & \text{if } pos(i) \neq \varnothing, \\
        \boldsymbol{z}_i^{\prime}, & \text{otherwise}
    \end{array}
    \right\}_{i \in \mathcal{B}}$ \Comment{by Eq.~(\ref{eq:pos_i})}

    \State Optimize model with $\mathcal{L}\left( \boldsymbol{Z}, \boldsymbol{Z}^{\prime}_{\text{SSPS}} \right)$

    \State Insert $\hat{\boldsymbol{Y}}$ into $\Qref$
    \State Insert $\boldsymbol{Z}^{\prime}$ into $\Qpos$
\end{algorithmic}
\end{algorithm}

\section{Experimental setup}
\label{sec:exp_setup}

\subsection{Datasets and feature extraction}

All models are trained on the VoxCeleb2 \cite{chung2018VoxCeleb2} dev set, which was collected "in the wild" from YouTube videos and contains \num{1092009} utterances from \num{5994} speakers distributed within \num{145569} videos. The evaluation is performed using VoxCeleb \cite{nagrani2017VoxCeleb} \textit{original} (VoxCeleb1-O), \textit{extended} (VoxCeleb1-E) and \textit{hard} (VoxCeleb1-H) trials. Speaker labels are discarded for the self-supervised training.

The duration of audio segments is \SI{2}{\second} by default. For DINO, local and global views correspond to four \SI{2}{\second} and two \SI{4}{\second} segments, respectively. The length of the reference segment used by the SSPS framework is \SI{4}{\second}. Audio segments are randomly sampled from an utterance and may overlap. Input features are obtained by computing $D_{\text{inp}}$-dimensional log-mel spectrograms using the torchaudio library, with $D_{\text{inp}}=40$ by default and $D_{\text{inp}}=80$ for DINO. The Hamming window length is \SI{25}{\milli\second}, and the frame-shift is set to \SI{10}{\milli\second}. Voice Activity Detection (VAD) is not applied as training samples consist mostly of continuous speech segments. Input features are normalized using instance normalization.

\subsection{Data-augmentation}

SSL commonly relies on extensive data-augmentation techniques to learn representations that are robust against extrinsic variabilities, such as environmental and channel noise, recording devices, and varying acoustic conditions.

Following other SSL approaches for SV \cite{zhang2021SimCLR, xia2021MoCo, lepage2022LabelEfficient}, two types of transformations are applied to the input signals at the beginning of each training iteration. First, reverberation is applied by randomly selecting an RIR from the simulated Room Impulse Response database \cite{ko2017StudyReverberantSpeechRobustSR}. Then, the MUSAN dataset \cite{snyder2015MUSAN} is used to randomly sample background noises, overlapping music tracks, or speech segments. To simulate a variety of real-world scenarios, the Signal-to-Noise Ratio (SNR) is randomly sampled between $\left[13; 20\right]$ dB for speech, $\left[5; 15\right]$ dB for music, and $\left[0; 15\right]$ dB for noises.

As the DINO framework benefits from a more diversified data-augmentation strategy, different augmentation conditions are applied: either no augmentation, reverberation, noise, or a combination of reverberation and noise.

\subsection{SSL frameworks}

The encoder $f$ is based on the ECAPA-TDNN \cite{desplanques2020ECAPATDNN} (22.5M parameters) architecture. For preliminary experiments, the encoder $f$ is replaced by the Fast ResNet-34 \cite{chung2020DefenceMetricLearningSR} (1.4M parameters) architecture, enabling faster experimental iteration for SSPS hyperparameter tuning. The Fast ResNet-34 has a base dimension of 16 and relies on Self-Attentive Pooling (SAP) to produce utterance-level representations. The ECAPA-TDNN has a hidden dimension set to 1024 and uses Attentive Statistics Pooling (ASP). Both encoders output representations of dimension $D_{\text{repr}}=512$, which are used to perform SV.

The projector $g$ is a standard MLP composed of three linear layers. A Batch Normalization and a ReLU activation function follow each layer except the last one. The hidden dimension is set to 2048, and the last layer outputs $D_{\text{emb}}=512$ units. The projector is discarded for contrastive methods (SimCLR and MoCo) as it degrades the downstream performance.

Models are trained for 100 epochs using Adam optimizer with no weight decay, a batch size of 256, and a learning rate set to 0.001, which is reduced by 5\% every 5 epochs. For DINO, the number of epochs is reduced to 80, the optimizer is SGD, the batch size is set to 128, the weight decay is set to $5e^{-5}$, and the learning rate is linearly ramped up to 0.2 during the 10-epochs warm-up before following a cosine scheduler from 0.2 to $1e^{-5}$.

The supervised baseline corresponds to an equivalent model trained with the AAM-Softmax loss ($s=30$, $m=0.2$) in a fully supervised way using the training set labels.

The code is available as part of the \textit{sslsv}\footnote{\url{https://github.com/theolepage/sslsv}} toolkit, based on the PyTorch framework. Trainings are performed on 2 $\times$ NVIDIA Tesla V100 16 GB and 4 $\times$ NVIDIA Tesla V100 32 GB for DINO.

\subsubsection{SimCLR} The temperature is set to $\tau=0.03$, and the loss implementation is symmetric. 
\subsubsection{MoCo} The temperature is set to $\tau=0.03$. The memory queue has a size of $|\boldsymbol{Q}^{\text{MoCo}}|=\num{65536}$ negatives. The coefficient of the EMA update is fixed to $m=0.999$.

\subsubsection{SwAV} The temperature is set to $\tau=0.1$ and the number of prototypes to $P=3000$. The queue is enabled after 15 epochs, and its size is set to $|\boldsymbol{Q}^{\text{SwAV}}|=\num{3840}$. The prototypes are frozen during the first epoch. The loss implementation is symmetric.

\subsubsection{VICReg} The scaling hyperparameters are set as follows: $\lambda=1$, $\mu=1$, and $\nu=0.04$, respectively, for the invariance, variance, and covariance terms.

\subsubsection{DINO} The projector output dimension is reduced to 256, and a final layer is introduced to map the $l_2$-normalized embeddings to $D_{\text{emb}}=\num{65536}$ units. This last layer has weight normalization and is frozen during the first epoch. The student and teacher temperatures are set to $\tau_{\text{s}}=0.1$ and $\tau_{\text{t}}=0.04$, respectively. The coefficient of the EMA update increases from $m=0.996$ to $m=1.0$ using a cosine scheduler. Gradients with a norm greater than 3.0 are clipped.

\subsection{SSPS}

Different positive sampling strategies are presented in the experiments: `SSL' corresponds to the default same-utterance sampling used in SSL frameworks, `SSPS' refers to the proposed approach, and `Supervised' represents a supervised positive sampling baseline that generates anchor-positive pairs from different recordings using the training set labels. These systems are trained using the corresponding positive sampling technique during 20 epochs starting from the end of the standard SSL training.

The SSPS algorithm is vectorized and processes mini-batches at each training iteration in Distributed Data-Parallel (DDP) mode. The size of the reference memory queue $\Qref$ is set to $|\Qref|=N$. The size of the positive memory queue $\Qpos$ is set to $|\Qpos|=N$ for SSPS-NN and to $|\Qpos|=K$ for SSPS-Clustering. K-means is performed at the beginning of every epoch using a custom PyTorch implementation running on the GPU for 10 iterations. Experiments are conducted on 2 $\times$ NVIDIA Tesla A100 80 GB and 4 $\times$ NVIDIA Tesla A100 80 GB for DINO.

The hyperparameter search is conducted using the SimCLR framework with the Fast ResNet-34 encoder, and the number of additional training epochs is reduced from 20 to 10. By default, the number of prototypes is set to $K=\num{25000}$, and the sampling window hyperparameter is set to $M=50$ and  $M=1$ for SSPS-NN and SSPS-Clustering, respectively.

Additional metrics are reported when SSPS is enabled by using the training set labels:
\begin{itemize}
    \item \textit{Speaker accuracy (\%)}: corresponds to the average number of pseudo-positive samples matching the anchor's speaker identity;
    \item \textit{Recording accuracy (\%)}: corresponds to the average number of pseudo-positive samples matching the anchor's recording (i.e., VoxCeleb video).
\end{itemize}
The objective of the proposed approach is to simultaneously maximize speaker accuracy while minimizing recording accuracy to encourage diversity in recording conditions across anchors and their corresponding pseudo-positives.

\subsection{Evaluation protocol}

To evaluate the performance of the systems on SV, the model weights of the last 10 epochs are averaged to provide more consistent results. Representations are extracted by processing full-length audio samples from the test set. The scoring of each trial is determined by computing the cosine similarity of $l_2$-normalized representation pairs.

Following NIST Speaker Recognition evaluation protocol \cite{sadjadi2017NISTSRE2016}, the performance is reported in terms of Equal Error Rate (EER) and minimum Detection Cost Function (minDCF) with $P_{target} = 0.01$, $C_{\text{miss}}=1$ and $C_{\text{fa}}=1$.
\section{Results and discussions}
\label{sec:results}

\subsection{Effect of the positive sampling on SSL performance}
\begin{table}[t]
  \titlecaption{Evaluation of SSL frameworks with SSL and Supervised positive sampling strategies}{The encoder is ECAPA-TDNN, and the EER is reported on VoxCeleb1-O. The relative EER reduction \textbf{$\Delta$ (\%)} shows the improvement when using Supervised over SSL positive sampling. Best results are in \textbf{bold}.}
  \label{tab:sslsv}
  \centering
  \begin{tabular}{llS[table-format=1.2]S[table-format=1.4]S[table-format=1.2]}
    \toprule    
    \multirow{2}{*}{\textbf{Framework}} & \multirow{2}{*}{\textbf{Pos. sampling}} & \multicolumn{2}{c}{\textbf{VoxCeleb1-O}} & {\multirow{2}{*}{\textbf{$\Delta$ (\%)}}} \\
    \cmidrule(lr){3-4}
    & & \textbf{EER (\%)} & \textbf{$\text{minDCF}_\text{0.01}$} & \\
    \midrule

    \multirow{2}{*}{SimCLR} & SSL & 6.30 & 0.5286 & {\multirow{2}{*}{72.62}} \\
     & Supervised & \bfseries 1.72 & \bfseries 0.2395 & \\
    \midrule

    \multirow{2}{*}{MoCo} & SSL & 6.20 & 0.5501 & {\multirow{2}{*}{71.77}} \\
     & Supervised & \bfseries 1.75 & \bfseries 0.2547 & \\
    \midrule

    \multirow{2}{*}{SwAV} & SSL & 7.97 & 0.6097 & {\multirow{2}{*}{45.06}} \\
     & Supervised & \bfseries 4.38 & \bfseries 0.4837 & \\
    \midrule

    \multirow{2}{*}{VICReg} & SSL & 7.70 & 0.5883 & {\multirow{2}{*}{41.30}} \\
     & Supervised & \bfseries 4.52 & \bfseries 0.4993 & \\
    \midrule

    \multirow{2}{*}{DINO} & SSL & 3.07 & 0.3616 & {\multirow{2}{*}{23.14}} \\
     & Supervised & \bfseries 2.36 & \bfseries 0.2712 & \\
     \midrule

     \rowcolor{BaselineRowBg} Supervised & & 1.34 & 0.1521 & \\
    \bottomrule
  \end{tabular}
\end{table}

The performance of SSL frameworks on SV (VoxCeleb1-O) with the ECAPA-TDNN encoder using both SSL and Supervised positive sampling strategies is reported in Table~\ref{tab:sslsv}.

Among the evaluated SSL frameworks, the state-of-the-art DINO method achieves the best performance on SV. The best three methods in terms of EER on VoxCeleb1-O are DINO (3.07\%), MoCo (6.20\%), and SimCLR (6.30\%).

SimCLR and MoCo provide competitive results considering their lower training complexity compared to DINO. This suggests that contrastive-based objective functions are particularly effective for SV applications. Among them, SimCLR stands out as a compelling compromise, balancing strong downstream performance with relatively low architectural complexity.

The supervised baseline achieves an EER of 1.34\% on VoxCeleb1-O, highlighting the performance gap that remains between SSL and supervised systems. This work hypothesizes that the primary bottleneck is the same-utterance positive sampling at the core of SSL frameworks. As demonstrated in Table~\ref{tab:sslsv}, significant performance improvements are observed when leveraging VoxCeleb metadata to create anchor-positive pairs from distinct recordings of the same speaker. Supervised positive sampling results in EER reductions of approximately 73\%, 72\%, 45\%, 41\%, and 23\% for SimCLR, MoCo, SwAV, VICReg, and DINO, respectively, on VoxCeleb1-O.

These findings reveal that the default SSL same-utterance positive sampling strategy is a significant limitation that hinders the performance of SSL frameworks. Adopting a different positive sampling method could significantly enhance the effectiveness of SSL frameworks on SV tasks.

The following methods are selected to represent each group of SSL framework paradigms in the subsequent experiments: SimCLR (contrastive learning), SwAV (clustering), VICReg (information maximization), and DINO (self-distillation).

\subsection{Hyperparameter tuning for SSPS}

\begin{table}[t]
  \scriptsize
  \titlecaption{Hyperparameters search for SSPS-NN and SSPS-Clustering in terms of SV performance}{The model is SimCLR, the encoder is Fast ResNet-34, and the benchmark is VoxCeleb1-O. Best results for each method are in \textbf{bold}.}
  \label{tab:ssps_hyperparams}
  \centering
  \begin{tabular}{lccS[table-format=2.2]S[table-format=1.4]}
    \toprule    
    \multirow{2}{*}{\textbf{Positive sampling}} & \multicolumn{2}{c}{\textbf{SSPS Hyperparameters}} & \multicolumn{2}{c}{\textbf{VoxCeleb1-O}} \\
    \cmidrule(lr){2-3} \cmidrule(lr){4-5}
    & $K$ & $M$ & \textbf{EER (\%)} & \textbf{$\text{minDCF}_\text{0.01}$} \\
    \midrule
    SSL & \multicolumn{1}{c}{} & \multicolumn{1}{c}{} & 9.41 & 0.6378 \\
    \midrule
    \multirow{5}{*}{SSPS-NN} & \multicolumn{1}{c}{} & 1 & 8.85 & 0.6240 \\
    & \multicolumn{1}{c}{} & 10 & 8.65 & 0.6186 \\
    & \multicolumn{1}{c}{} & 25 & 8.38 & 0.6189 \\
    & \multicolumn{1}{c}{} & 50 & \bfseries 8.15 & \bfseries 0.6145 \\
    & \multicolumn{1}{c}{} & 100 & 8.41 & 0.6164 \\
    \midrule
    \multirow{25}{*}{SSPS-Clustering} & \num{6000} & 0 & 6.63 & 0.5493 \\
    \cmidrule(lr){2-5} & \num{10000} & 0 & 6.82 & 0.5629 \\
    \cmidrule(lr){2-5} & \multirow{5}{*}{\num{25000}} & 0 & 7.30 & 0.5805 \\
    & & 1 & 5.80 & \bfseries 0.5250 \\
    & & 2 & \bfseries 5.73 & 0.5258 \\
    & & 3 & 6.06 & 0.5410 \\
    & & 5 & 6.85 & 0.5672 \\
    \cmidrule(lr){2-5} & \multirow{5}{*}{\num{50000}} & 0 & 7.75 & 0.5829 \\
    & & 1 & 6.27 & 0.5351 \\
    & & 2 & \bfseries 6.11 & 0.5394 \\
    & & 3 & 6.12 & 0.5343 \\
    & & 5 & 6.15 & \bfseries 0.5282 \\
    \cmidrule(lr){2-5} & \multirow{5}{*}{\num{100000}} & 0 & 8.25 & 0.6046 \\
    & & 1 & 7.07 & 0.5725 \\
    & & 2 & 6.87 & \bfseries 0.5549 \\
    & & 3 & 6.72 & 0.5570 \\
    & & 5 & \bfseries 6.58 & 0.5627 \\
    \cmidrule(lr){2-5} & \multirow{5}{*}{\num{150000}} & 0 & 8.29 & 0.6170 \\
    & & 1 & 7.54 & 0.5923 \\
    & & 2 & 7.13 & 0.5711 \\
    & & 3 & 7.06 & 0.5766 \\
    & & 5 & \bfseries 6.92 & \bfseries 0.5483 \\
    \midrule
    \multirow{2}{*}{SSPS-Clustering (C)} & \num{6000} & 0 & 13.31 & 0.8125 \\
    \cmidrule(lr){2-5} & \num{25000} & 1 & \bfseries 11.04 & \bfseries 0.7453 \\
    \midrule
    \rowcolor{BaselineRowBg} Supervised & & & 3.93 & 0.3900 \\
    \bottomrule
  \end{tabular}
\end{table}

\begin{figure}[t]
  \centering
  \subfloat[\textbf{SSPS-NN}\label{fig:ssps_nn_hyperparams}]{
    \includegraphics[width=0.45\linewidth]{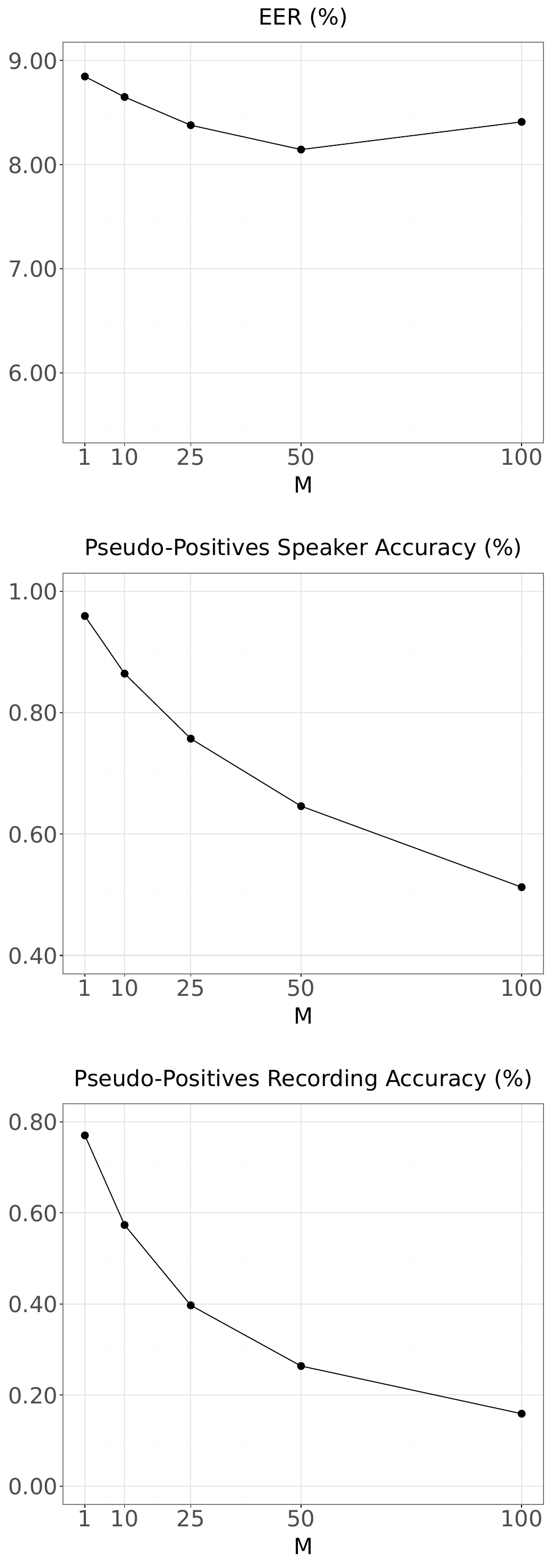}
  }
  \hfill
  \subfloat[\textbf{SSPS-Clustering}\label{fig:ssps_clustering_hyperparams}]{
    \includegraphics[width=0.45\linewidth]{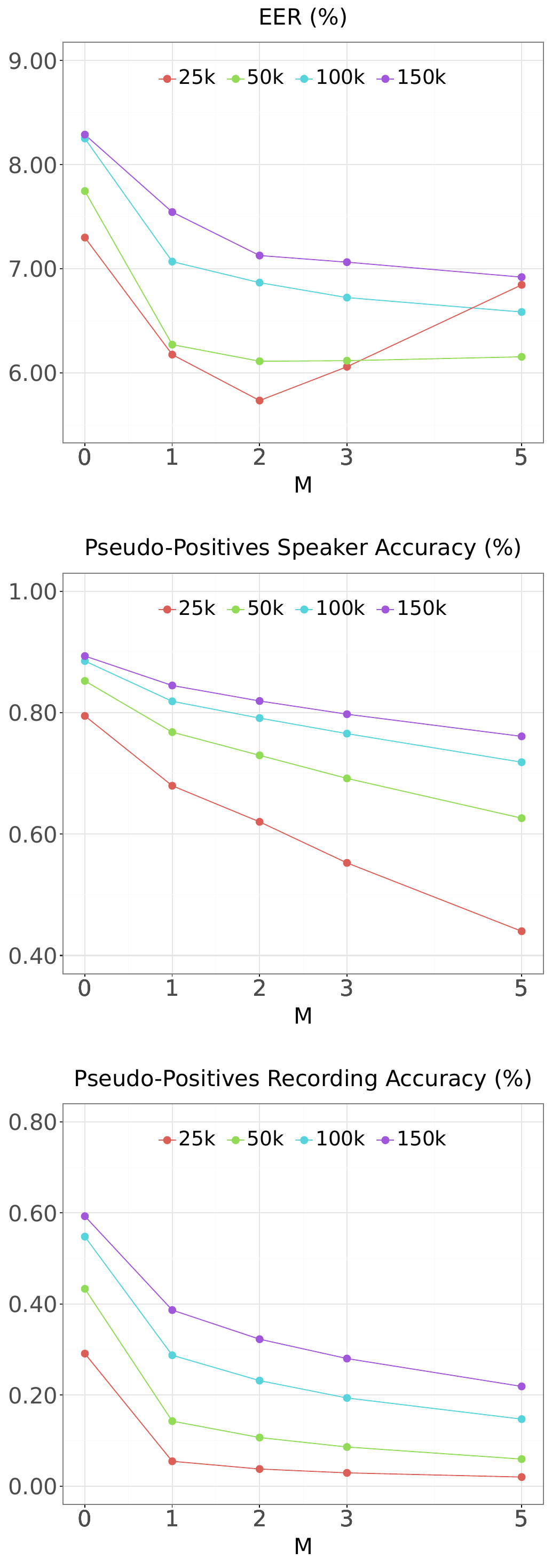}
  }
  \titlecaption{Hyperparameters search for SSPS-NN (a) and SSPS-Clustering (b) in terms of EER, pseudo-positive speaker accuracy, and pseudo-positive recording accuracy}{The model is SimCLR, the encoder is Fast ResNet-34, and the EER is reported on VoxCeleb1-O.}
  \label{fig:ssps_hyperparams}
\end{figure}

In this work, it is hypothesized that the primary bottleneck of SSL frameworks lies in the way positive samples are selected. The preliminary results of SSPS on SV (VoxCeleb1-O), obtained with SimCLR and the Fast ResNet-34 encoder, with different sampling algorithms and hyperparameters, are reported in Table~\ref{tab:ssps_hyperparams}. These results are compared with SSL and Supervised positive sampling baselines, achieving 9.41\% EER and 3.93\% EER on VoxCeleb1-O. Additionally, Figure~\ref{fig:ssps_hyperparams} illustrates the effect of $K$ and $M$ in terms of EER on VoxCeleb1-O, as well as pseudo-positives speaker accuracy and pseudo-positives recording accuracy. 

\subsubsection{SSPS-NN}
SSPS-NN samples a pseudo-positive close to the anchor in the latent space with different sampling window sizes $M$. The optimal configuration is achieved with $M=50$, which implies that the algorithm should sample pseudo-positives close to the anchor to ensure they belong to the same speaker identity, while still maintaining a large sampling window to identify pseudo-positives from different recording sources. There is a clear trade-off between pseudo-positive speaker and pseudo-positive recording accuracies, as a small sampling window can prevent the method from selecting different recordings, as highlighted in Figure~\subref*{fig:ssps_nn_hyperparams}. With $M=50$, SSPS-NN effectively improves SV performance, providing pseudo-positives with a speaker accuracy of approximately 50\% and a recording accuracy of approximately 25\%.

\subsubsection{SSPS-Clustering}
SSPS-Clustering samples a pseudo-positive according to the clustering assignments of the latent representations. By using $K=\num{6000}$, which approximately corresponds to the number of speakers in the training set, and sampling from the same cluster as the anchor with $M=0$, the EER is reduced to 6.63\%. Setting $K=\num{25000}$ and $M=1$ further improves the EER to 5.80\%. Although the lowest EER is obtained with $K=\num{25000}$ and $M=2$, this configuration is not considered, because the setting with $M=1$ corresponds to the best configuration in terms of minDCF. Setting $K=\num{150000}$, which represents the actual number of recordings in the training set, results in worse performance, which may be caused by the fact that several recordings may be included in the same cluster. As shown in Figure~\subref*{fig:ssps_clustering_hyperparams}, by using a large value for $K$ and sampling from a neighboring cluster (i.e., $M > 0$), the overall performance is better, and the pseudo-positive speaker accuracy is maintained while the pseudo-positive recording accuracy drops compared to SSPS-NN. Note that SSPS-Clustering (C), using the sampling cluster centroid as the pseudo-positive embedding, proves ineffective, suggesting that the embedding space is dense but not continuous. SSPS-Clustering is selected for the following as it allows sampling pseudo-positives with higher speaker accuracy and lower recording accuracy compared to SSPS-NN.

\subsection{Evaluation of SSPS on SV}

\begin{table*}[t]
  \titlecaption{Evaluation of SSL frameworks on SV with different positive sampling strategies: SSL, SSPS, and Supervised}{The encoder is ECAPA-TDNN, and the benchmarks are VoxCeleb1-O, VoxCeleb1-E, and VoxCeleb1-H. The best methods are highlighted in \colorbox{SelectedRowBg}{light blue} and the baselines are highlighted in \colorbox{BaselineRowBg}{light gray}. Best results are in \textbf{bold}, excluding supervised baselines.}
  \label{tab:ssps}
  \centering
  \begin{tabular}{llccS[table-format=1.2]S[table-format=1.4]S[table-format=1.2]S[table-format=1.4]S[table-format=2.2]S[table-format=1.4]}
    \toprule    
    \multirow{2}{*}{\textbf{Framework}} & \multirow{2}{*}{\textbf{Pos. sampling}} & \multicolumn{2}{c}{\textbf{SSPS Hyperparameters}} & \multicolumn{2}{c}{\textbf{VoxCeleb1-O}} & \multicolumn{2}{c}{\textbf{VoxCeleb1-E}} & \multicolumn{2}{c}{\textbf{VoxCeleb1-H}} \\
    \cmidrule(lr){3-4} \cmidrule(lr){5-6} \cmidrule(lr){7-8} \cmidrule(lr){9-10}
    & & $K$ & $M$ & \textbf{EER (\%)} & \textbf{$\text{minDCF}_\text{0.01}$} & \textbf{EER (\%)} & \textbf{$\text{minDCF}_\text{0.01}$} & \textbf{EER (\%)} & \textbf{$\text{minDCF}_\text{0.01}$} \\
    \midrule

        \multirow{4}{*}{SimCLR} & SSL & & & 6.30 & 0.5286 & 6.86 & 0.5599 & 10.98 & 0.6692 \\
        & SSPS & \num{6000} & 0 & 2.90 & 0.3206 & 3.38 & 0.3292 & 6.13 & 0.4887 \\
        & \cellcolor{SelectedRowBg}SSPS & \cellcolor{SelectedRowBg}\num{25000} & \cellcolor{SelectedRowBg}1 & \cellcolor{SelectedRowBg} \bfseries 2.57 & \cellcolor{SelectedRowBg} \bfseries 0.3033 & \cellcolor{SelectedRowBg} \bfseries 3.11 & \cellcolor{SelectedRowBg} \bfseries 0.3125 & \cellcolor{SelectedRowBg} \bfseries 5.56 & \cellcolor{SelectedRowBg} \bfseries 0.4638 \\
        & \cellcolor{BaselineRowBg}Supervised & \cellcolor{BaselineRowBg} & \cellcolor{BaselineRowBg} & \cellcolor{BaselineRowBg} 1.72 & \cellcolor{BaselineRowBg} 0.2395 & \cellcolor{BaselineRowBg} 1.88 & \cellcolor{BaselineRowBg} 0.2314 & \cellcolor{BaselineRowBg} 3.66 & \cellcolor{BaselineRowBg} 0.3641 \\
        \midrule
    
        \multirow{4}{*}{SwAV} & SSL & & & 7.97 & 0.6097 & 8.87 & 0.7052 & 15.15 & 0.8273 \\
        & SSPS & \num{6000} & 0 & 7.07 & 0.5847 & 7.96 & 0.6803 & 13.93 & 0.8149 \\
        & \cellcolor{SelectedRowBg}SSPS & \cellcolor{SelectedRowBg}\num{25000} & \cellcolor{SelectedRowBg}1 & \cellcolor{SelectedRowBg} \bfseries 6.50 & \cellcolor{SelectedRowBg} \bfseries 0.5687 & \cellcolor{SelectedRowBg} \bfseries 7.35 & \cellcolor{SelectedRowBg} \bfseries 0.6507 & \cellcolor{SelectedRowBg} \bfseries 13.03 & \cellcolor{SelectedRowBg} \bfseries 0.8014 \\
        & \cellcolor{BaselineRowBg}Supervised & \cellcolor{BaselineRowBg} & \cellcolor{BaselineRowBg} & \cellcolor{BaselineRowBg} 4.38 & \cellcolor{BaselineRowBg} 0.4837 & \cellcolor{BaselineRowBg} 4.92 & \cellcolor{BaselineRowBg} 0.5538 & \cellcolor{BaselineRowBg} 9.49 & \cellcolor{BaselineRowBg} 0.7254 \\
        \midrule
    
        \multirow{4}{*}{VICReg} & SSL & & & 7.70 & 0.5883 & 9.05 & 0.7170 & 15.25 & 0.8431 \\
        & SSPS & \num{6000} & 0 & 7.45 & 0.5513 & 8.56 & 0.6926 & 14.15 & 0.8343 \\
        & \cellcolor{SelectedRowBg}SSPS & \cellcolor{SelectedRowBg}\num{25000} & \cellcolor{SelectedRowBg}1 & \cellcolor{SelectedRowBg} \bfseries 6.95 & \cellcolor{SelectedRowBg} \bfseries 0.5262 & \cellcolor{SelectedRowBg} \bfseries 8.16 & \cellcolor{SelectedRowBg} \bfseries 0.6759 & \cellcolor{SelectedRowBg} \bfseries 13.51 & \cellcolor{SelectedRowBg} \bfseries 0.8263 \\
        & \cellcolor{BaselineRowBg}Supervised & \cellcolor{BaselineRowBg} & \cellcolor{BaselineRowBg} & \cellcolor{BaselineRowBg} 4.52 & \cellcolor{BaselineRowBg} 0.4993 & \cellcolor{BaselineRowBg} 5.43 & \cellcolor{BaselineRowBg} 0.6343 & \cellcolor{BaselineRowBg} 10.37 & \cellcolor{BaselineRowBg} 0.7908 \\
        \midrule
    
        \multirow{4}{*}{DINO} & SSL & & & 3.07 & 0.3616 & 3.32 & 0.4003 & 6.20 & 0.5731 \\
        & SSPS & \num{6000} & 0 & 2.78 & 0.3140 & 3.07 & 0.3456 & 5.66 & 0.5158 \\
        & \cellcolor{SelectedRowBg}SSPS & \cellcolor{SelectedRowBg}\num{25000} & \cellcolor{SelectedRowBg}1 & \cellcolor{SelectedRowBg} \bfseries 2.53 & \cellcolor{SelectedRowBg} \bfseries 0.2843 & \cellcolor{SelectedRowBg} \bfseries 2.55 & \cellcolor{SelectedRowBg} \bfseries 0.3150 & \cellcolor{SelectedRowBg} \bfseries 4.93 & \cellcolor{SelectedRowBg} \bfseries 0.4632 \\
        & \cellcolor{BaselineRowBg}Supervised & \cellcolor{BaselineRowBg} & \cellcolor{BaselineRowBg} & \cellcolor{BaselineRowBg} 2.36 & \cellcolor{BaselineRowBg} 0.2712 & \cellcolor{BaselineRowBg} 2.41 & \cellcolor{BaselineRowBg} 0.2986 & \cellcolor{BaselineRowBg} 4.64 & \cellcolor{BaselineRowBg} 0.4378 \\
        \midrule
    
        \rowcolor{BaselineRowBg} Supervised & & & & 1.34 & 0.1521 & 1.49 & 0.1736 & 2.84 & 0.2887  \\
        \bottomrule
  \end{tabular}
\end{table*}

The performance of SSL frameworks on SV with different configurations of the SSPS-Clustering strategy is reported using the ECAPA-TDNN encoder in Table~\ref{tab:ssps}.

Configurations using $K=\num{25000}$ and $M=1$ outperform those using $K=\num{6000}$ and $M=0$, which confirms the assumption that speaker representations are grouped according to channel information in latent space. In the following, SSPS-Clustering is referred to as SSPS with the former set of hyperparameters, as it corresponds to the best performance on the downstream task.

Compared to SSL positive sampling, which serves as the baseline, SSPS improves the EER and minDCF of all reported SSL frameworks across all VoxCeleb benchmarks. This shows the effectiveness of the proposed approach when applied to various SSL paradigms. As expected, the best result is achieved by DINO with SSPS, yielding 2.53\% EER and 0.2843\% minDCF on VoxCeleb1-O. Note that this performance is very close to its Supervised positive sampling baseline, resulting in 2.36\% EER and 0.2712 minDCF. Furthermore, SimCLR provides comparable performance to DINO by reaching 2.57\% EER and 0.3033 minDCF. The performance of SwAV and VICReg is also improved with SSPS, from 7.97\% to 6.50\% EER and from 7.70\% to 6.95\% EER, respectively. The lower performance improvement on SwAV and VICReg can be attributed to the already limited performance of their Supervised positive sampling baselines, suggesting that the same-utterance positive sampling is not the only bottleneck. Therefore, SSPS proves to be very effective in improving the downstream performance of major SSL frameworks on SV. For comparison, the supervised method achieves 1.34\% EER and 0.1521 minDCF on VoxCeleb1-O.

SimCLR, based on the contrastive learning paradigm, demonstrates a remarkable improvement over its baseline when using SSPS, as it results in a 58\% relative EER reduction. This is very promising, as the system outperforms DINO without SSPS, despite DINO relying on a far more complex training framework. Supervised positive sampling strategy applied to SimCLR achieves 1.72\% EER and 0.2395 minDCF, implying that there is potential for further improvements of SSL contrastive-based methods.

Finally, it can be observed that: (1) DINO with SSPS reaches a performance comparable to its Supervised positive sampling baseline, suggesting that this approach has reached its full potential; (2) SimCLR with Supervised positive sampling results in the best performance among SSL methods, outperforming DINO with Supervised positive sampling. This shows that the focus should be on SSL contrastive frameworks, which benefit from the discriminative capacity of the contrastive-based objective function. Therefore, the following experiments will be conducted with the SimCLR framework.

\subsection{Comparison to state-of-the-art SSL methods for SV}

\begin{table}[t]
  \titlecaption{Comparison of state-of-the-art SSL methods for SV}{Methods are grouped into contrastive and self-distillation frameworks. The benchmark is VoxCeleb1-O, with EERs sorted in descending order. The proposed methods are in \textbf{bold} and highlighted in \colorbox{SelectedRowBg}{light blue}. Other results are drawn from the literature.}
  \label{tab:eval}
  \centering
  \begin{tabular}{lS[table-format=2.2]S[table-format=1.4]}
    \toprule    
    \multirow{2}{*}{\textbf{Method}} & \multicolumn{2}{c}{\textbf{VoxCeleb1-O}} \\
    \cmidrule(lr){2-3}
    & \textbf{EER (\%)} & \textbf{$\text{minDCF}_\text{0.01}$} \\
    \midrule

    \multicolumn{3}{c}{\textit{Contrastive}} \\
    \cmidrule(){1-3}
    AP + AAT \cite{huh2020AAT} & 8.65 & \\
    Contrastive + VICReg \cite{lepage2022LabelEfficient} & 8.47 & 0.6400 \\
    SimCLR + MSE loss \cite{zhang2021SimCLR} & 8.28 & 0.6100 \\
    MoCo + ProtoNCE \cite{xia2021MoCo} & 8.23 & 0.5900 \\
    CEL \cite{mun2020CEL} & 8.01 & \\
    SimCLR + AAT \cite{tao2022LGL} & 7.36 & \\
    C3-MoCo \cite{zhang2022C3DINO} & 6.40 & \\
    DPP \cite{tao2023DPP} & 2.89 & \\
    \rowcolor{SelectedRowBg} \textbf{SimCLR-SSPS} & \bfseries 2.57 & \bfseries 0.3033 \\
    \midrule

    \multicolumn{3}{c}{\textit{Self-distillation}} \\
    \cmidrule(){1-3}
    SSReg \cite{sang2022SSReg} & 6.99 & \\
    DINO + Cosine loss \cite{han2022DLGLC} & 6.16 & 0.5240 \\
    DINO \cite{cho2022DINO} & 4.83 & 0.4630 \\
    DINO + Curriculum \cite{heo2022DINOCurriculum} & 4.47 & \\
    CA-DINO \cite{han2024CADINO} & 3.59 & 0.3529  \\
    RDINO \cite{chen2023RDINO} & 3.29 & \\
    MeMo \cite{jin2024WGVKT} & 3.10 & \\
    RDINO + W-GVKT \cite{jin2024MeMo} & 2.89 & 0.3330 \\
    DINO + RMP \cite{kim2024RMP} & 2.62 & \\
    \rowcolor{SelectedRowBg} \textbf{DINO-SSPS} & \bfseries 2.53 & \bfseries 0.2843 \\
    DINO + Aug. \cite{chen2022ComprehensiveStudySelfDistillation} & 2.51 & \\
    C3-DINO \cite{zhang2022C3DINO} & 2.50 & \\
    \bottomrule
  \end{tabular}
\end{table}

A comparative evaluation of recent state-of-the-art SSL methods for SV on the VoxCeleb1-O benchmark is presented in Table~\ref{tab:eval}. The methods are organized into contrastive and self-distillation frameworks.

Among contrastive methods, the proposed SimCLR-SSPS outperforms all prior approaches and sets a new state-of-the-art of 2.57\% EER on the benchmark. This confirms the effectiveness of the bootstrapped positive sampling strategy within SSL contrastive frameworks. Although DPP \cite{tao2023DPP} achieves a comparable EER of 2.89\%, its positive sampling leverages audio-visual cues from speech-face pairs of VoxCeleb2, whereas SSPS is trained using audio data only.

In the self-distillation category, the proposed DINO-SSPS also demonstrates competitive performance compared to the other methods listed by reaching 2.53\% EER. While DINO + Aug. \cite{chen2022ComprehensiveStudySelfDistillation} and C3-DINO \cite{zhang2022C3DINO} achieve marginally lower EERs (2.51\% and 2.50\%, respectively), their performance is comparable to DINO-SSPS, given the minimal difference. Notably, the former relies on a more complex data-augmentation strategy, while the latter initializes the DINO model from pre-trained MoCo weights. In contrast, DINO-SSPS achieves competitive results without these specific enhancements.

These results validate the effectiveness of the proposed positive sampling strategy when integrated into both contrastive and self-distillation SSL frameworks, as it explicitly addresses a key limitation of existing approaches.

\subsection{Effect of SSPS on SSL speaker representations}

\begin{figure}[t]
  \centering
  \subfloat[\textbf{Test data} (VoxCeleb1)\label{fig:intra_speaker_sim_vox1}]{
    \centering
    \includegraphics[width=0.49\linewidth]{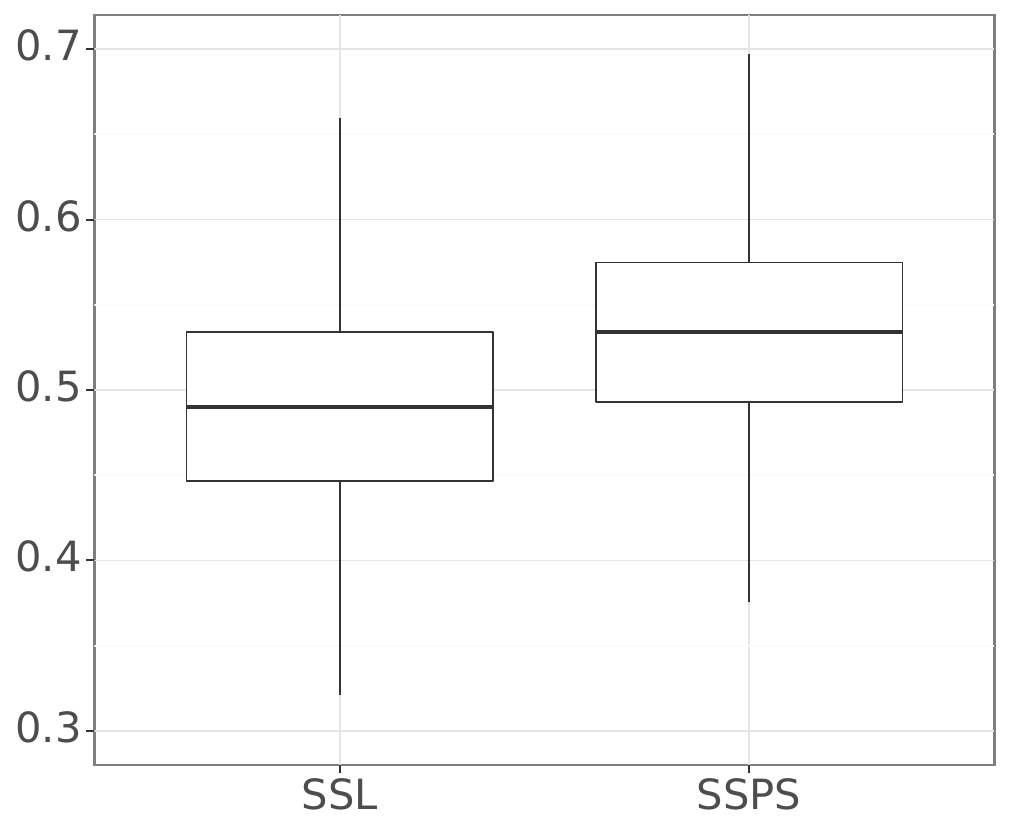}
  }
  \subfloat[\textbf{Train data} (VoxCeleb2)\label{fig:intra_speaker_sim_vox2}]{
    \centering
    \includegraphics[width=0.49\linewidth]{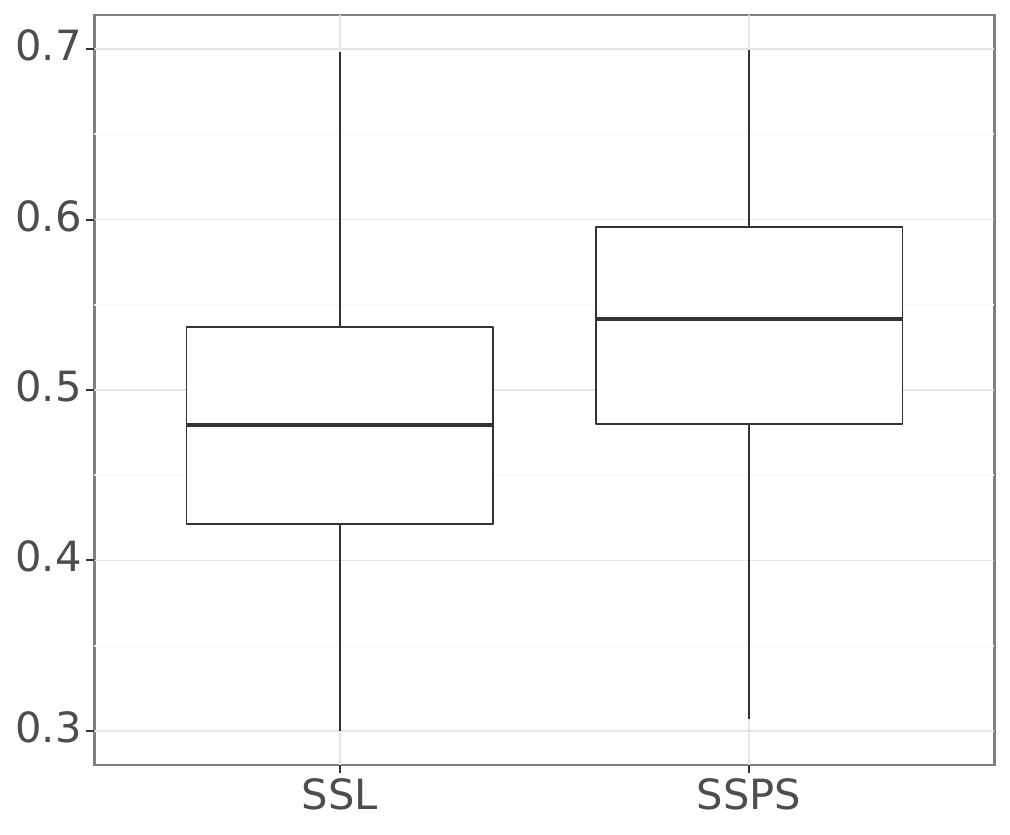}
  }
  \titlecaption{Intra-speaker cosine similarity on test (a) and train (b) samples for SSL and SSPS}{The model is SimCLR, and the encoder is ECAPA-TDNN.}
  \label{fig:intra_speaker_sim}
\end{figure}

\begin{figure}[t]
  \centering
  \subfloat[\textbf{SSL}\label{fig:tsne_baseline}]{
    \centering
    \includegraphics[width=0.49\linewidth]{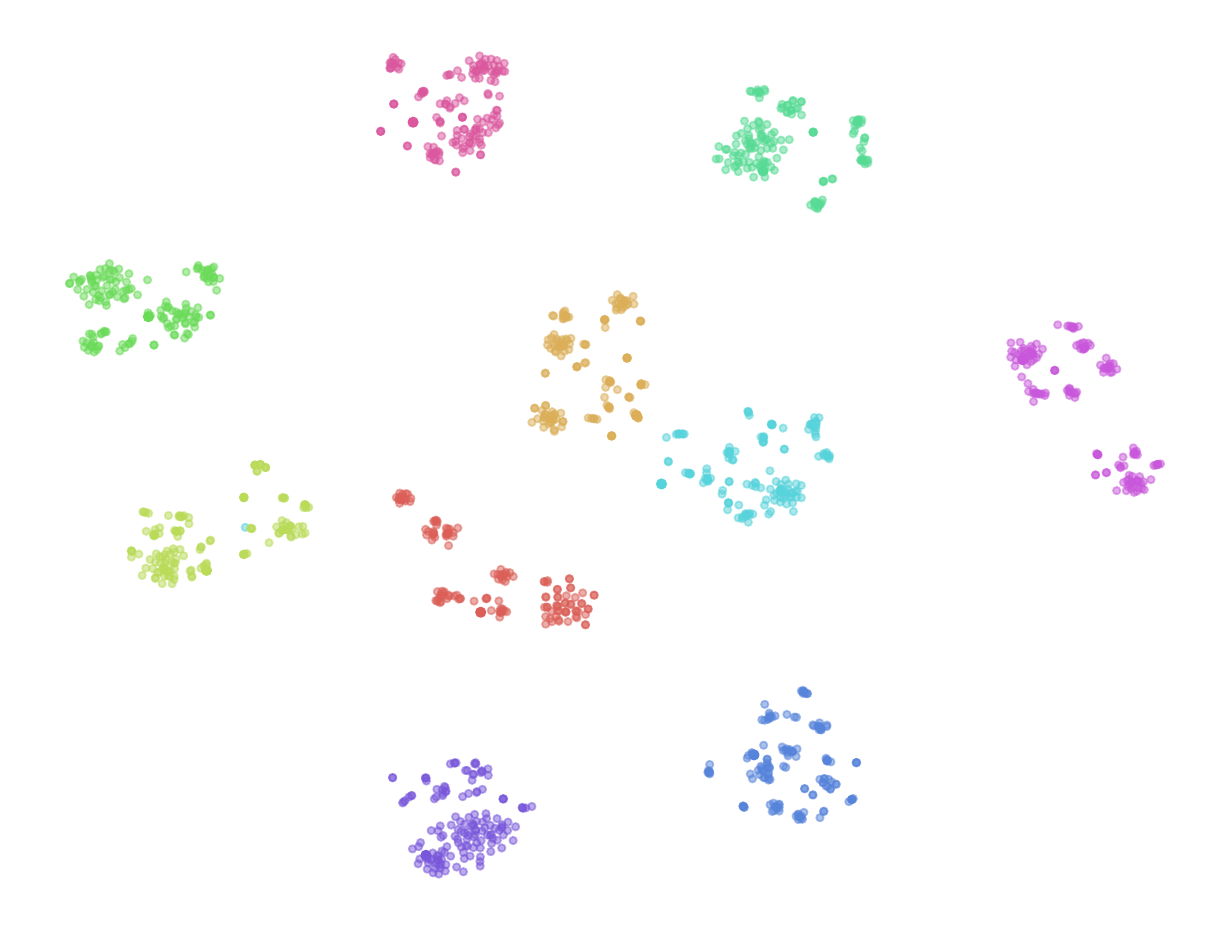}
  }
  \subfloat[\textbf{SSPS}\label{fig:tsne_ssps}]{
    \centering
    \includegraphics[width=0.49\linewidth]{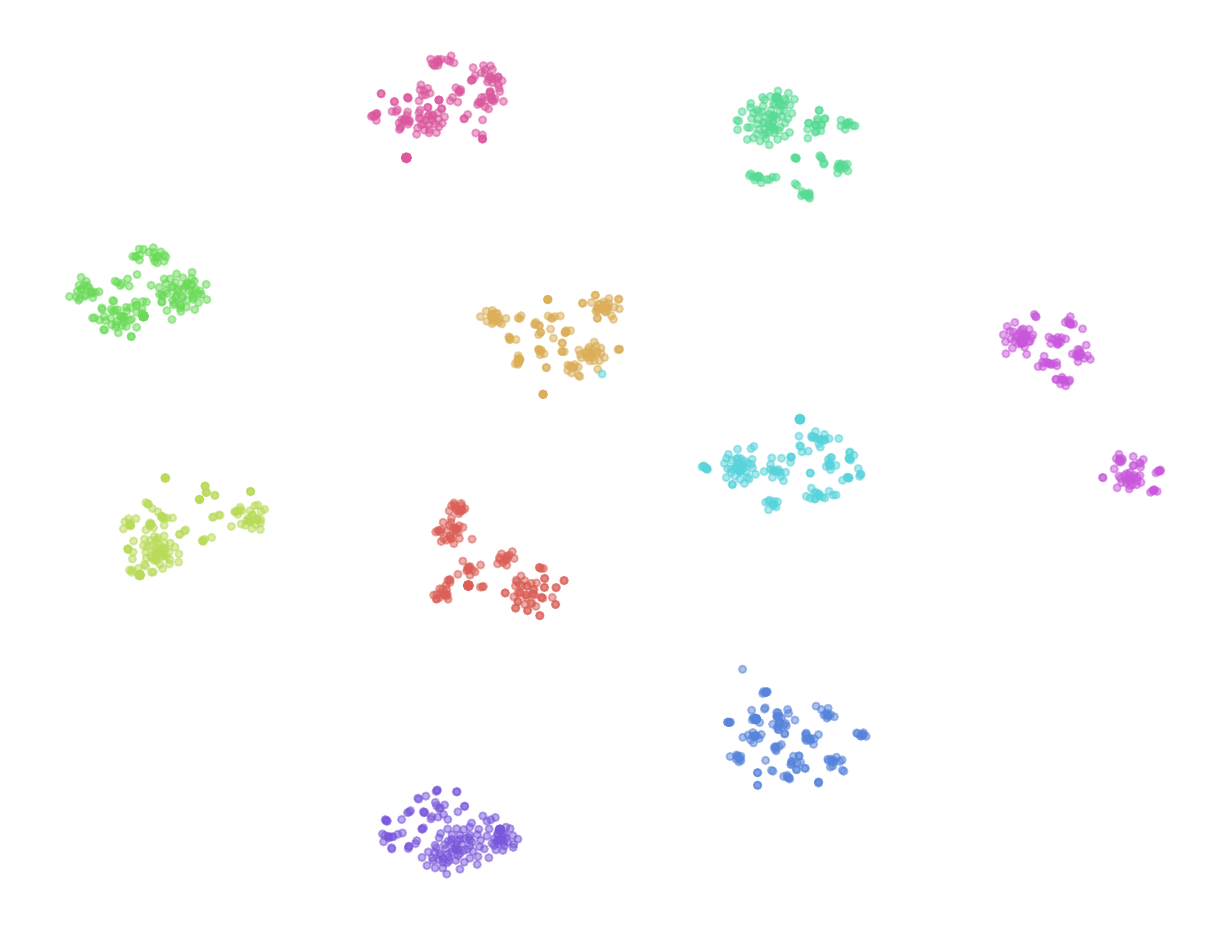}
  }
  \titlecaption{Visualization of speaker representations with SSL (a) and SSPS (b)}{The t-SNE is performed on representations from 10 speakers of VoxCeleb1. The model is SimCLR, and the encoder is ECAPA-TDNN.}
  \label{fig:tsne}
\end{figure}

The median cosine similarity of same-speaker representations for train and test samples increases when using SSPS compared to SSL positive sampling, as shown in Figure~\ref{fig:intra_speaker_sim}. This effect was expected, as providing more diverse positives to the SSL framework allows matching different representations with distinct channel characteristics to the same speaker identity, thereby reducing intra-class variability. This improvement in class compactness is visible in  Figure~\ref{fig:tsne}, which represents the t-SNE of 10 speakers' representations from VoxCeleb1 with SSL and SSPS positive sampling techniques. Some groups of samples (``sub-clusters'') remain distant in the latent space despite belonging to the same speaker identity, suggesting that SSPS may not effectively consider all utterances from a given speaker as pseudo-positives. This demonstrates a limitation of the proposed approach whenever the SSL model has already separated same-speaker representations in the latent space, due to strong acoustic variability, prior to the activation of SSPS.

\begin{figure}[t]
    \centering
    \includegraphics[width=0.9\linewidth]{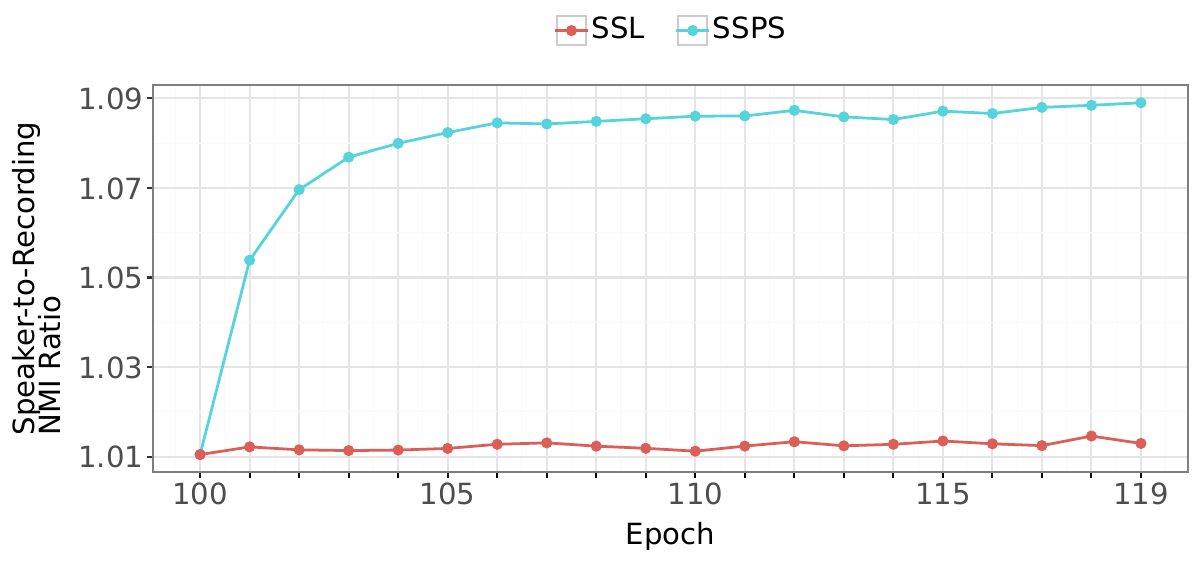}
    \titlecaption{NMI ratio of speaker and recording information with SSL and SSPS across training epochs}{The model is SimCLR, the encoder is ECAPA-TDNN, and the NMI is computed with $K=\num{6000}$ clusters on reference representations extracted from VoxCeleb2.}
    \label{fig:nmi}
\end{figure}

To further assess the quality of the speaker representations learned with SSPS, the Normalized Mutual Information (NMI)~\cite{strehl2002ClusterEnsembles} between the clustering assignments of the reference representations, obtained with $K=\num{6000}$, and the VoxCeleb2 metadata, including speaker and recording labels, is used to quantify the alignment with the underlying structure of the data. The NMI between two sets of assignments $U$ and $V$ is defined as:
\begin{equation}
    \text{NMI}(U, V) = \frac{2 \cdot I(U; V)}{H(U) + H(V)},
\end{equation}
where $I(U; V)$ is the Mutual Information (MI) between $U$ and $V$, $H(U)$ is the entropy of $U$, and $H(V)$ is the entropy of $V$. The NMI should be maximized, as it reflects the degree of alignment between the predicted cluster assignments and the ground-truth labels. This allows reporting the ratio of speaker-to-recording NMI across epochs with SSL and SSPS positive sampling strategies in Figure~\ref{fig:nmi}. Before the activation of SSPS, the NMI ratio of $\sim$1.01 implies that the representations contain as much information on speaker identity as on the recording source. At the end of the training, SSPS has increased the ratio of speaker over recording NMI to $\sim$1.09, which represents a $\sim$8\% relative improvement, while the SSL baseline results in a ratio close to the initial value. Thus, SSPS enables the model to learn representations that are more robust to extrinsic variabilities arising from different recording conditions by encoding less channel information compared to SSL same-utterance positive sampling.

\subsection{Robustness of SSPS to the absence of data-augmentation}

\definecolor{Red}{HTML}{eb3838}
\definecolor{Green}{HTML}{21bf52}
\newcommand{\cmark}{\textcolor{Green}{\ding{51}}}
\newcommand{\xmark}{\textcolor{Red}{\ding{55}}}

\begin{table}[t]
  \titlecaption{Effect of data-augmentation on SV performance with SSL and SSPS}{The model is SimCLR, the encoder is ECAPA-TDNN, and the benchmark is VoxCeleb1-O. Best results are in \textbf{bold}.}
  \label{tab:data-aug}
  \centering
  \begin{tabular}{lccS[table-format=2.2]S[table-format=1.4]}
    \toprule
    \multirow{2}{*}{\textbf{Positive sampling}} & \multirow{2}{*}{\textbf{Data-aug.}} & \multicolumn{2}{c}{\textbf{VoxCeleb1-O}} \\
    \cmidrule(lr){3-4}
    & & \textbf{EER (\%)} & \textbf{$\text{minDCF}_\text{0.01}$} \\
    \midrule
    \multirow{2}{*}{SSL}  & \cmark & \bfseries 6.30  & \bfseries 0.5286 \\
                          & \xmark & 15.00 & 0.7575 \\
    \midrule
    \multirow{2}{*}{SSPS} & \cmark & \bfseries 2.57 & 0.3033 \\
                          & \xmark & 2.77 & \bfseries 0.2840 \\
    \bottomrule
  \end{tabular}
\end{table}

\begin{figure}
    \centering
    \includegraphics[width=0.9\linewidth]{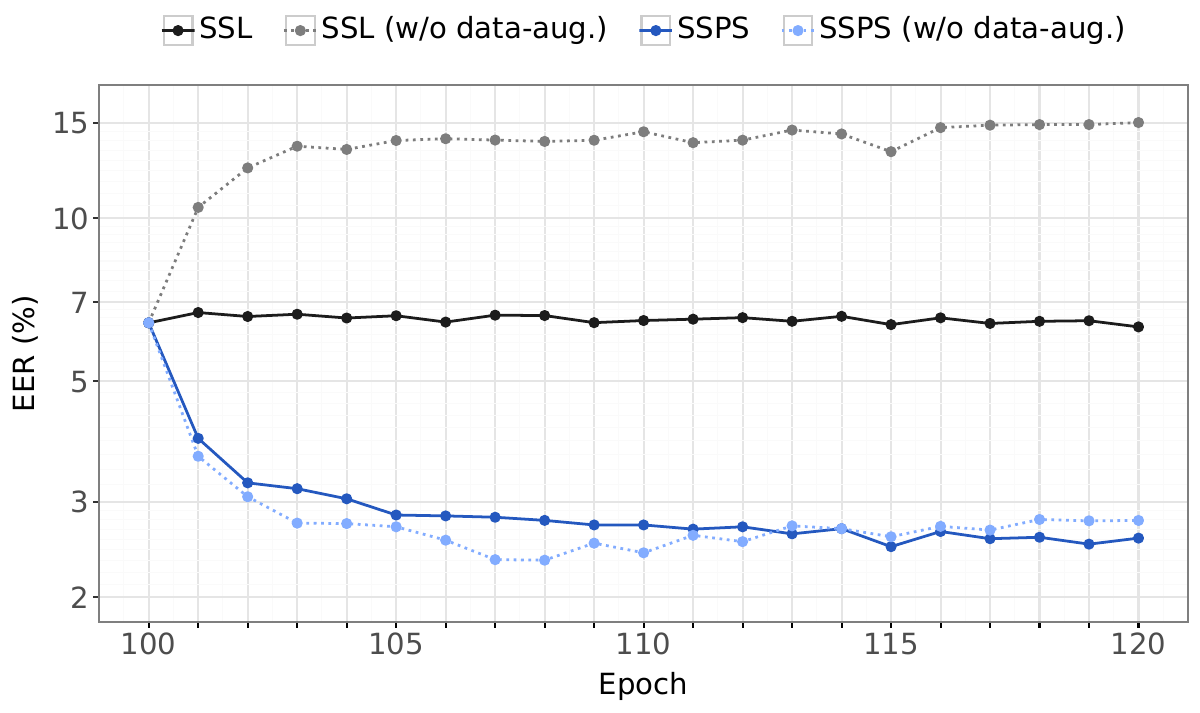}
    \titlecaption{Training convergence on SV of SSL and SSPS with and without data-augmentation}{The model is SimCLR, the encoder is ECAPA-TDNN, and the EER is reported on VoxCeleb1-O.}
    \label{fig:data_aug}
\end{figure}

SSL frameworks heavily depend on data-augmentation to construct meaningful anchor-positive pairs, as it introduces necessary variability to prevent the model from collapsing and learning trivial representations. However, since the VoxCeleb dataset consists of samples collected "in the wild," it inherently captures a wide range of natural acoustic conditions. This reduces the need for data-augmentation when an effective positive sampling strategy is implemented.

This experiment evaluates the ability of the proposed approach to converge without any data-augmentation pre-processing steps by comparing SSL and SSPS positive sampling techniques with their counterparts that do not involve data-augmentation. The results reported with SimCLR (ECAPA-TDNN) in Table~\ref{tab:data-aug} show that, without data-augmentation, the performance of the baseline significantly degrades from 6.30\% to 15.00\% EER on VoxCeleb1-O, whereas SSPS maintains strong performance with only a minor degradation of the EER from 2.57\% to 2.77\%, demonstrating its robustness to the absence of data-augmentation. Notably, the minDCF of SSPS even improves slightly when no data-augmentation is employed, decreasing from 0.3033 to 0.2840. This highlights the effectiveness of SSPS, as it outperforms the same-utterance positive sampling used in SSL, while simultaneously eliminating the need to artificially create diverse acoustic conditions.

The EER on VoxCeleb1-O across epochs for these systems is shown in Figure~\ref{fig:data_aug}. The model based on SSL positive sampling without data-augmentation (light gray, dashed) suffers from the similarity of channel characteristics between the anchor and the positive, leading to performance degradation from the first epoch. In contrast, the model based on SSPS maintains consistent performance, with (blue, solid) and without (light blue, dashed) data-augmentation, throughout training. This figure highlights the strong reliance of the standard same-utterance positive sampling on data-augmentation. This justifies the underlying intuition of SSPS, which is to create anchor-positive pairs from different recordings of the same speaker, rather than relying on augmentation techniques.

The reduced reliance of the proposed method on data-augmentation is an important property, considering that data-augmentation is a fundamental component for training SSL frameworks. However, data-augmentation is not an ideal approach as it depends heavily on the downstream task and input data, and often involves a hyperparameter space too large to be explored efficiently in many scenarios. This suggests that alternative combinations of positive sampling and data-augmentation strategies should be investigated.
\section{Conclusions}
\label{sec:conclusions}

To overcome the limitations of the default SSL same-utterance positive sampling, Self-Supervised Positive Sampling (SSPS) is proposed as a new strategy for sampling positives in SSL frameworks. For SV, this new approach identifies relevant and diverse pseudo-positives in the latent space that are not derived from the same recording as the anchor, which reduces the channel and recording information in speaker representations. SSPS enhances downstream performance for all major SSL frameworks, including SimCLR, SwAV, VICReg, and DINO, on VoxCeleb benchmarks compared to the default positive sampling approach. The best performance is achieved with SimCLR and DINO, reaching 2.57\% and 2.53\% EER, respectively, on VoxCeleb1-O, where the supervised baseline achieves 1.34\% EER. Notably, SimCLR with SSPS yields a 58\% relative reduction in EER. When using Supervised positive sampling, better performance is achieved with SimCLR compared to DINO, significantly reducing the gap with the fully supervised baseline, which motivates the need to consider contrastive-based SSL frameworks for the task of SV. Furthermore, better class compactness and a reduction of channel and recording information from the speaker representations are observed. Finally, the proposed method is shown to be more robust to the absence of data-augmentation than standard SSL models. This work addresses key limitations of SSL methods for SV, advancing the field toward closing the performance gap between supervised and self-supervised systems.

\section*{Acknowledgements}

This work was performed using HPC resources from GENCI-IDRIS (Grant 2023-AD011014623) and has been partially funded by the French National Research Agency (project APATE - ANR-22-CE39-0016-05).

\bibliography{biblio,biblio_additional}

\end{document}